\documentclass[traditabstract]{aa} 
\usepackage{graphicx}
\usepackage{txfonts}
\usepackage{natbib}

\usepackage{lscape}
\bibpunct{(}{)}{;}{a}{}{,} 

\usepackage{morefloats}

\usepackage{afterpage}

\usepackage{ulem}

\graphicspath{{Figures/}}

\usepackage{xspace}
\usepackage{color}
\usepackage{hyperref}

\newcommand{\logg}{\mbox{$\log g$~}}
\newcommand{\Tef}{\mbox{\,$T_{\rm eff}$~}}

\newcommand{\Vt}{\mbox{$V_t$~}}
\newcommand{\de}{de$^{\mathrm{a}}$}
\begin{document}

\title{Validating post-AGB candidates in the LMC and SMC using SALT\thanks{Based on observations made with the Southern African Large Telescope (SALT)}  spectra}

  
   \author{R. Szczerba\inst{1}
                  \and 
          M. Hajduk\inst{2}
          \and 
          Ya.V. Pavlenko\inst{3}
          \and 
          B.J. Hrivnak\inst{4}
          \and           
          B.M. Kaminsky\inst{3}
          \and 
          K. Volk\inst{5}
          \and 
          N. Si{\'o}dmiak\inst{1}
          \and
          I. Gezer\inst{1},\inst{6}
          \and \\
          L. Za\v{c}s\inst{7}
          \and
          W. Pych\inst{8}
          \and
          M. Sarna\inst{8}
          }

   \institute{Nicolaus Copernicus Astronomical Center,
   Rabia{\'n}ska 8, 87-100 Toru{\'n}, Poland
   \and
   Space Radio-Diagnostics Research Centre, University of Warmia and Mazury in Olsztyn, 10-719 Olsztyn, Poland 
   \and 
   Main Astronomical Observatory, NASU, Zabolotnoho 27, 03680 Kyiv, Ukraine
   \and
   Department of Physics and Astronomy, Valparaiso University, Valparaiso, IN 46383, USA
   \and 
   Space Telescope Science Institute, 3700 San Martin Drive, Baltimore, MD, 21218, USA
    \and
   Warsaw University Astronomical Observatory, Al. Ujazdowskie 4, 00-478, Warszawa, Poland
    \and
   Laser Center, Faculty of Physics and Mathematics, University of Latvia, Raiņa bulvaris 19, LV-1586 Riga, Latvia 
   \and
   Nicolaus Copernicus Astronomical Center, Bartycka 18, 00-716 Warszawa, Poland}

\mail{szczerba@ncac.torun.pl}
\authorrunning{Szczerba et al.}
\titlerunning{SALT observations of Post-AGB candidates in MCs}
   \date{Received ; accepted }

  \abstract
{We selected a sample of post-AGB candidates in the Magellanic Clouds on the 
basis of their near- and mid-infrared colour characteristics. Fifteen of the most optically 
bright  post-AGB candidates were observed with the South African Large Telescope in order to 
determine their stellar parameters and thus to validate or discriminate their nature as post-AGB objects 
in the Magellanic Clouds.
The spectral types of
absorption-line objects 
were estimated according to the MK classification, and effective temperatures were obtained by means of  
stellar atmosphere modelling. Emission-line objects 
were classified on the basis of the 
fluxes of the emission lines and the presence of the continuum. Out of 15 observed objects, only 4 appear to be genuine post-AGB stars (27\%). 
In the SMC, 1 out of 4 is post-AGB, and in the LMC, 3 out 11 are post-AGB objects. 
Thus, we can conclude that the selected region in the colour-colour diagram, while selecting 
the genuine post-AGB objects, overlaps severely with other types of objects, in particular young stellar objects and planetary nebulae. Additional classification criteria are required to distinguish between post-AGB stars and other types of objects. In particular, photometry at far-IR wavelengths would greatly assist in distinguishing young stellar objects from evolved ones. On the other hand, we showed that the low-resolution optical spectra appear to be sufficient to determine whether the candidates are post-AGB objects. }

 
\maketitle

\section{Introduction}

The stars with the initial mass range of about $\rm 1-8\,M_{\odot}$ lose most 
of their mass during the asymptotic giant branch (AGB) phase \citep[see e.g.][for review]{Habing:2004aa}. Once the star
leaves the AGB, it starts to increase its surface temperature and the mass loss gradually decreases. The star is heavily obscured by an expanding dusty
envelope until the envelope disperses into the interstellar medium.
Post-AGB\footnote{Sometimes they are referred to as proto-planetary nebulae (PPN).
However, those post-AGB stars, which evolve too slowly, may avoid the PN phase.} stars comprise a
group of objects in the transition stage between the AGB and planetary nebula
(PN) phase. Most of them are quite faint or even undetectable in the optical. Their
spectral energy distribution (SED) is dominated by a dusty shell, which is bright in the infrared. The SED show one or two distinguishable peaks (one  corresponding
to the shell and the other  to the central star, if visible), depending on
the orientation and characteristics of the dusty shell
\citep[e.g.][]{2008ApJ...677..382S}. 
Only about 300 
post-AGB stars are known in our Galaxy. 
This small number is due primarily 
to their short evolutionary timescale \citep{2007A&A...469..799S}.

The Magellanic Clouds (MCs) provide an excellent environment to study stellar populations
thanks to the known distances and negligible extinction. 
However, only about 
70 optically bright post-AGB candidates have been suggested in the Large
Magellanic Cloud (LMC) \citep{2011A&A...530A..90V} and 21 high-probability
post-AGB candidates have been found in Small Magellanic Cloud (SMC)
\citep{2014MNRAS.tmp..343K} using optical spectroscopy. \cite{Kamath:2015aa} extended the analysis of  \cite{2011A&A...530A..90V}, and selected 35 likely post-AGBs in the LMC on the basis of determined stellar parameters.

We selected a sample of post-AGB candidates in the MCs using solely 
the infrared colours (for details, see Sect.\ref{sec:selection}) of known post-AGB stars from our Galaxy and the MCs. The selected region on the colour-colour diagram (CCD) is not free from other kinds of objects, namely
planetary nebulae, young stellar objects (YSOs), or Seyfert galaxies. 
While these objects have different spectral characteristics in the
optical than post-AGB stars, they could have similar photometric colours or magnitudes. 
However, Seyfert galaxies can be easily 
distinguished by means
of the radial velocity shift, while PNe show highly excited emission lines requiring
a hot central source.

\begin{figure*}
\resizebox{\hsize}{!}{\includegraphics{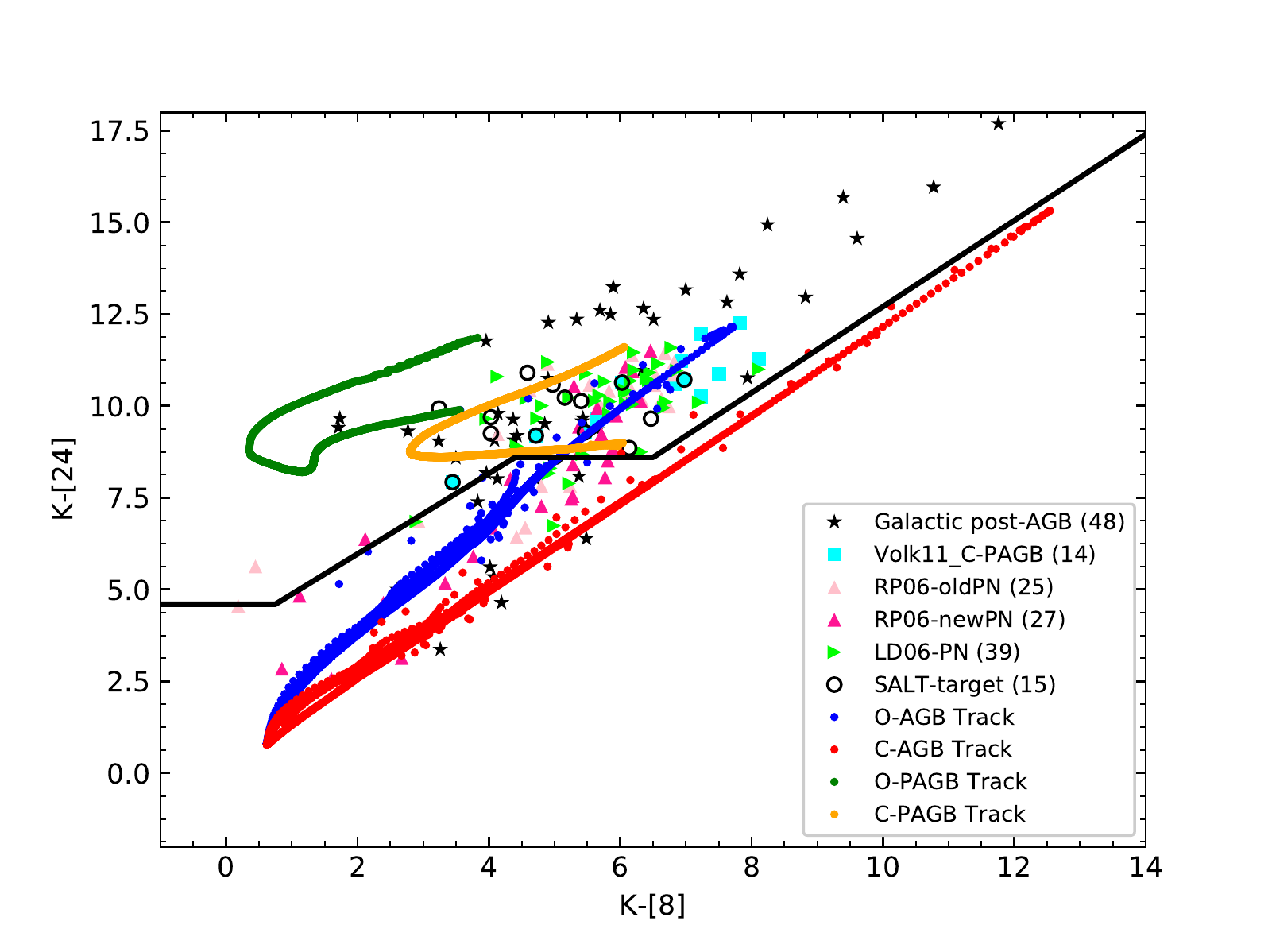}}
\caption{Colour-colour diagram used for the selection of the post-AGB candidates. The main post-AGB candidate area is above the black lines. The numbers in parentheses 
in the inset list the total number of each class of objects plotted in this figure (see text for details).}
\label{pagb-select}
\end{figure*}

We performed spectroscopic observations of the brightest objects in
optical wavelengths. The sample was selected on the basis of the optical photometry from
\citet{2004AJ....128.1606Z} for the LMC and \citet{2002AJ....123..855Z} for the
SMC. In one case, data from the NOMAD catalogue \citep{Zacharias:2004aa} were used.
The immediate goal is to know which objects in  the sample belong to the small class of post-AGB objects so that we might add to the confirmed sample and better understand 
their properties, while the overall goal is 
to determine the fraction of post-AGB stars and other types of objects (planetary nebulae, 
young stellar objects) that are present in the infrared-selected sample.

The paper is organised as follows. In Section\,2 we describe our criteria used for the selection of post-AGB candidates in the MCs, and in Section\,3 we describe briefly their South African Large Telescope (SALT) observations and present complementary photometric data. In  Section 4 we discuss results, starting with emission-line objects and then  the objects with absorption lines in their spectra. Model atmospheres are used to determine physical parameters of the latter group. 
Finally, we present the discussion and summary.

\section{Selection of  post-AGB candidates}
\label{sec:selection}

\citet{2011A&A...530A..90V}  identified some high-probability post-AGB objects in  the LMC, using a combination of infrared photometry from the Spitzer Space Telescope \citep[SST:][]{Werner:2004aa} and ground-based optical data. The work on selecting post-AGB candidates in the LMC was then extended by \cite{Kamath:2015aa}.
A similar work to search for optically visible post-AGB candidates in the SMC was carried out by \citet{2014MNRAS.tmp..343K}.
Interestingly, post-red giant branch (post-RGB) stars were recognised in the SMC and LMC by \citet{2014MNRAS.tmp..343K} and \cite{Kamath:2015aa}, respectively. In addition, the authors   show  that SEDs indicative of discs (results of binary evolution) are dominant among post-AGB and post-RGB 
objects.

In all these works spectroscopic confirmation (determination of the stellar parameters from the spectral observations) was still necessary to recognise the real nature of the object.
However, the large initial number of candidates
required a correspondingly large number of spectra to be obtained. For example, \citet{2014MNRAS.tmp..343K} obtained spectra for 801 SMC objects to find out that 21 of them are `likely post-AGB' and 42 are post-RGB objects (about 8\%), while in the case of the LMC \cite{Kamath:2015aa} classified 35 sources as `likely post-AGB', and 119  as post-RGB objects by analysis spectra for 2102 sources (i.e. about 7\% are confirmed post-AGB and/or RGB objects).

In general, there are no good mid-IR colour-colour or colour-magnitude criteria for selecting post-AGB candidates. The AGB stars, which generally have  simpler geometry of circumstellar shells than post-AGB objects, can be more easily recognised on CCDs or colour-magnitude diagrams (CMDs) \citep[see e.g.][]{Blum:2006aa, Sewio:2013aa, Matsuura:2014ve}. For example, the CMDs discussed by \cite{Blum:2006aa} are shown for LMC objects  from \citet{Jones:2017aa} in Appendix\,\ref{appB}. While some groups of objects (like AGBs, massive stars, or planetary nebulae) are located in relatively distinct regions of the CCDs, the post-AGB objects are highly mixed with these groups (especially with YSOs and PNe).

To select post-AGB candidates in the Magellanic Clouds, we   used K$-[8]$ versus K$-[24]$ CCD.
This diagram can be used for sources with and without known distances. We verified that among different CCDs constructed from near-IR and SST 
photometry, this CCD is the best suited for separating evolved stars from other types of objects \citep{Szczerba:2016aa, Matsuura:2014ve, Jones:2017aa}.
Both O- and C-rich AGB stars are nicely separated on this diagram and there is a 
region where post-AGB objects are located as well, but mixed especially with young stellar objects \citep{Szczerba:2016aa,Jones:2017aa}. The colour 
indices constructed using wavelengths longer than 24 $\mu$m would serve better as a discriminator  between evolved  and young stars. The reason is that very often YSOs are embedded in their parental cloud and this results in a larger far-IR excess than is observed from post-AGB objects. However, IRAS\,{\citep{Neugebauer:1984aa}, AKARI\,\citep{Murakami:2007aa}, and SST observations at longer wavelengths are not as numerous for the MCs objects as are the SST photometric data at 24 $\mu$m.
For 
the post-AGB candidates in the LMC and SMC on this CCD, we  used K photometry from 2MASS and 8 and 24 $\mu$m SST photometry from the Surveying the Agents of a Galaxy's Evolution (SAGE) project \citep{Meixner:2006aa} for the LMC and  SAGE-SMC \citep{Gordon:2010aa}.  

The first step was to  identify the region on this CCD  which contains most of the known post-AGB objects. For this purpose we  used a sample of Galactic post-AGB objects from the Torun catalogue 
\citep{2007A&A...469..799S, Szczerba:2012kx}.
In Figure \ref{pagb-select} we present their distribution,
based on synthetic photometry from their Infrared Space Observatory (ISO) spectra, as black star symbols (Galactic post-AGB in the inset on Fig. 1). The sample of post-AGB candidates in the LMC and SMC, known to us at the time of the SALT proposal preparation, from \citet{2011ApJ...735..127V}, are plotted as 
cyan squares (Volk11\_C-PAGB). The selection of post-AGB candidates in \cite{2011ApJ...735..127V} was based on the SAGE-Spec Spitzer Legacy Program \citep{Kemper:2010aa}. Using triangles of different colours and orientations we  show the positions of newly discovered (RP06-newPN) and previously known (RP06-oldPN) LMC planetary nebulae from \citet{Reid:2006aa}, as well as those from \citet{Leisy:2006aa} (LD06-PN). The number of sources plotted, limited by available photometry of good quality, is shown in parentheses in the inset.

In addition, in Figure \ref{pagb-select} we  overplot positions of  self-consistent time-dependent hydrodynamical (HD) radiative transfer calculations for gaseous dusty circumstellar shells around C-rich and O-rich stars in the final stages of their AGB/post-AGB evolution. These calculations are based on 
the evolutionary track for a star with an initial
mass of 3.0\,M$_{\odot}$, which, due to mass loss, is reduced to a typical mass for central stars of post-AGB objects of 0.605\,M$_{\odot}$ \citep{Bloecker:1995fk} at the end of the AGB evolution. We used a single dust  grain size (a=0.05 $\mu$m) for both chemical compositions: amorphous carbon (AC) and astronomical silicates (ASil) \citep[see][for details] {Steffen:1998aa}. The HD evolutionary track for AC dust is shown by red/orange dots (C-AGB Track/C-PAGB Track in the inset on Fig.1), while for ASil dust by blue/green dots (O-AGB Track/O-PAGB Track) for the AGB/post-AGB phase, respectively. 
The HD models are plotted in equidistant intervals in time (100 years for AGB and 0.3 year for post-AGB), so the density of the points is a direct measure of the probability of finding various objects  in different parts of the diagram. The last 200 years during AGB (the quick transition to post-AGB phase) are plotted each 20 years. The AGB phase covers 350\,000 years, while the post-AGB only about 1\,000 years. While new models for the evolution of post-AGB stars exist \citep{MillerBertolami2016}, their much faster evolution  results in optically thicker circumstellar envelopes for the post-AGB models, thus resulting in a smaller loops in the diagram. For the purpose of post-AGB candidates selection, the older evolutionary tracks of \cite{Bloecker:1995fk} seem to be sufficient and appropriate.

Taking into account the distribution of different classes of sources as well as the position of post-AGB HD tracks, we  defined somewhat arbitrarily a region which should contain post-AGB objects. This region is located in the upper left portion of Figure 1.
The boundary points, 
which allow us to reconstruct the lines delineating the lower boundary of the  
post-AGB region in the 
(K-$[8.0]$, K-$[24]$) plane are (-1.0, 4.6), (0.75, 4.6), (4.4, 8.6), (6.5, 8.6), and (14.0, 17.5). Although these boundary lines were based on post-AGB objects from the Galaxy, 
for the purpose of selecting post-AGB candidates in the two MCs,
we assume that they lie in the same region on the colour-colour plane. 
We note  that K-$[8.0]$ minimum and maximum values are arbitrary, and are related to the size of the plot. The selected  objects are distributed within $0<$K-$[8.0]<12$ (see below).

Altogether we   selected 716 and 274 post-AGB candidates in the LMC
and SMC, respectively, in this way.
However, this approach does not guarantee that only post-AGB objects are selected.   We should expect contamination by other sources with similar SEDs in the 2--24 $\mu$m range, like oxygen-rich AGB stars, PNe, and  especially YSOs \citep[see e.g.][]{Szczerba:2016aa}. To test our selection criteria for post-AGB candidates, we   selected a small subsample of 15 post-AGB candidates (SALT-target in the inset on Fig.\,1) in order to validate their nature with the SALT telescope. 
 We selected the brightest post-AGB candidates (the faintest has 16.6 mag in V).
For these candidates we expected to obtain $\rm S/N \geq 30$, which would be sufficient for   spectral type classification.
In addition, we observed for comparison purposes one relatively  bright star with a spectral type similar to that expected for the post-AGB stars, the yellow supergiant Sk 105. The observed sources are listed in Table\,\ref{objlist}.  This table contains  for each source its running number  (SK\,105 does not have an attributed  running number); SAGE designation, short name, and 2\,MASS designation (and IRAS and other names if they exist); and coordinates (sources are ordered according to right ascension,    R.A.) taken from the SAGE catalogues. Four of our post-AGB candidates come from the SMC, while 11 are from the LMC. 

\section{Observations}

\begin{table*}
\caption{Designations and positions for the standard star and post-AGB candidates in the Magellanic Clouds.}
\label{objlist}

\centering

{\tiny
\begin{tabular}{rlllrlllll}
\hline
\hline
No. & SAGE Name                      & Short Name & 2MASS Name                  & IRAS Name  &  Other Names     & R.A. (J2000)   & Dec. (J2000) \\ 
    &                                       &                     &                                          &                           &                            &   (deg)            &   (deg)            \\ \hline
     
  &   \object{J010243.04$-$720726.0}   & Sk 105     & J01024303$-$7207260    & 
          &  Sk 105                     & 15.679338
        & -72.123913 \\

1 & \object{J004747.63$-$731727.7}   & J004747       & J00474753$-$7317277        &                           & [M2002] SMC 9251 &  11.948482   &   -73.291038 \\
2 & \object{J004841.88$-$732615.2}   &  J004841      & J00484187$-$7326151    &                           &   LHA 115-N 31      & 12.174516     &    -73.437567   \\
3 & \object{J010546.42$-$714705.2}   &  J010546      & J01054645$-$7147053    &                           &   [BSS2007] 255    & 16.443433     &    -71.784802     \\
4 & \object{J011542.87$-$730959.3}   &  J011542      & J01154286$-$7309592    &                           &   LHA 115-N 86      & 18.928606     &    -73.166479     \\
5 & \object{J045747.93$-$662844.9}   & J045747       & J04574795$-$6628448    &                           &  MSX LMC 1225   & 74.449714     &   -66.479165  \\
6 & \object{J045907.37$-$654313.4}   & J045907       & J04590738$-$6543134    & 04589$-$6547   &  LHA 120-S 10     & 74.780710     &   -65.720396   \\
7 & \object{J051110.61$-$661253.8}  & J051110        & J05111065$-$6612537       & 05110$-$6616   & MSX LMC 287     &  77.794406    &    -66.214932  \\
8 & \object{J051228.18$-$690755.7}   & J051228       & J05122821$-$6907556    & 05127$-$6911   &  [WSI2008] 365  &  78.117368    &   -69.132202  \\
9 & \object{J052043.86$-$692341.0}   & J052043     & J05204385$-$6923403  & S05211$-$6926   & [WSI2008] 531 &  80.182631   &    -69.394729 \\
10 & \object{J052229.14$-$710814.9}   & J052229       & J05222918$-$7108149    & 05232$-$7111   &                            &  80.621451    &   -71.137499  \\
11 & \object{J052520.76$-$705007.5}   & J052520       & J05252077$-$7050075       & Z05259$-$7052 &                            &  81.336610    &   -70.835424  \\
12 & \object{J052915.66$-$673247.4}  & J052915       & J05291566$-$6732477   &                              & LHA 120-N 53 &  82.315290    &    -67.546480 \\
13 & \object{J053348.91$-$701323.6}   &  J053348     & J05334893$-$7013234        &                             & MSX LMC 755 &  83.453744    &    -70.223224 \\   
14 & \object{J054055.81$-$691614.6}   & J054055       & J05405584$-$6916146       &                           & [WSI2008] 1044 &  85.232873    &    -69.270746 \\
15 & \object{J055825.96$-$694425.8}   & J055825       & J05582596$-$6944257       & 05588$-$6944   & MSX LMC 1601 &  89.608238    &    -69.740510 \\                                                                                                                                   
\hline
\end{tabular}
}

\begin{list}{}{}
\item Note: Positions are from the SAGE LMC and SMC surveys for all the sources.
\end{list}

\end{table*}
\begin{table}
\caption{Log of spectroscopic observations.
} 
\label{log}
\centering
\begin{tabular}{ llllll}
\hline
\hline
date       & No.$^1$ & short    & set$^3$ & V$^4$ & exposure \\
           &         & name$^2$ &         &  [mag]        & time [s]\\ \hline
2011/09/30 &  4  & J011542 & red & 16.15 & 900$\times$2 \\
               &    & Sk 105  & red & 11.82 & 80$\times$2\\
               &    &         &     &       & 90$\times$2\\           
           &    &         &     &       & 180$\times$2  \\
2011/10/05 &  2 & J004841 & red & 15.40 & 1200  \\
           &    &        &     &        & 587   \\
                   & 15 & J055825 & red & 15.42 & 350$\times$2  \\
2011/10/09 & 6  & J045907 & red & 13.30 & 250$\times$2  \\
                   & 5  & J045747 & red & 14.82 & 310$\times$2  \\
                   & 7  & J051110 & red & 16.62 & 1100$\times$2 \\
2011/10/11 & 1  & J004747 & red & 15.38 & 400$\times$2  \\
                   & 14 & J054055 & red & 14.38 & 200$\times$2  \\
2011/10/28 & 15 & J055825 & blue & 15.42 &920$\times$2  \\
2011/11/14 & 3  & J010546 & red & 15.44 &320$\times$2   \\
2011/11/22 & 12 & J052915 & red & 16.63 & 920$\times$2  \\
2011/12/17 & 13 & J053348 & red & 15.56 & 510$\times$2  \\
2012/01/06 & 11 & J052520 & red & 15.34 &300$\times$2   \\
                   & 10 & J052229 & red & 16.58 &900$\times$2   \\
2012/01/07 & 11 & J052520 & blue & 15.34 &1500$\times$2 \\
2012/01/10 & 5  & J045747 & red & 14.82 &310$\times$2   \\
2012/01/11 & 9  & J052043 & red & 15.08 &300$\times$2   \\
                   & 7  & J051110 & red & 16.62 &1100$\times$2  \\
2012/01/29 & 8  & J051228 & red & 16.15 &900$\times$2 \\        
2012/02/29 & 6  & J045907 & red & 13.30 &150$\times$2   \\
\hline
\end{tabular}
\begin{list}{}{}
\item $^1$ See Tab.\ref{objlist} for the object's running number. \\
$^2$ See Tab.\ref{objlist} for the designations of the observed objects.\\
$^3$ setting\\
$^4$ See Tab.\ref{photometry} for the source of V photometry. 
\end{list}
\end{table}

\begin{table*}
\caption{Photometry for the standard star and the Magellanic Clouds post-AGB candidates.}
\label{photometry}

\centering

{\tiny
\begin{tabular}{rrrrrrrrrrrrrr}
\hline
\hline
 No. & Short Name  &  U       & B           &   V         & B-V & J          & H          &  K & [3.6]  &  [4.5]  & [5.8]   & [8.0]  & [24]  \\ \hline
 &  Sk 105     &  11.849   & 11.933      &  11.820      &  0.113 & 10.832     & 10.638     & 10.577 & 10.447 & 10.347    & 10.372  & 10.353 & 10.520 \\ 
 & $\pm$       &   0.076   &  0.163       &   0.300   &   &  0.023     &  0.025     &  0.025     &  0.036 & 0.039 &  0.031  &  0.0 & 0.137 \\    
1 &  J004747     &  14.53   & 15.43       &  15.38  & 0.05    & 15.555     & 15.447     & 15.610$^a$ & 15.496 &         & 14.126  & 12.371 & 5.676 \\ 
 & $\pm$       &   0.03   &  0.02     &    0.04 &      &  0.069     &  0.113     &  0.123     &  0.070 &         &  0.143  &  0.094 & 0.012 \\       
 2 & J004841     &  15.346  & 15.471      &  15.400 & 0.071    & 15.360     & 15.203     & 14.491     &        & 13.155  & 11.455  &  9.905 & 3.589 \\ 
 & $\pm$       &   0.151  &  0.086      &   0.073   &  &  0.056     &  0.108     &  0.101     &        &  0.044  &  0.038  &  0.067 & 0.006 \\ 
 3 & J010546     &  15.435  & 15.816      &  15.439 & 0.377    & 14.945     & 14.598     & 14.608     & 13.511 & 12.932  & 10.657  &  8.142 & 4.954 \\ 
 & $\pm$       &   0.032  &  0.029      &   0.057 &    &  0.058     &  0.105     &  0.111     &  0.044 &  0.042  &  0.042  &  0.029 & 0.011 \\       
 4 & J011542     &  16.453  & 17.282      &  16.150 & 1.132     & 14.793     & 14.295     & 13.603     & 12.242 & 11.559  & 10.425  &  8.200 & 3.469 \\ 
 & $\pm$       &   0.114  &  0.087      &   0.104  &   &  0.046     &  0.051     &  0.051     &  0.120 &  0.037  &  0.063  &  0.027 & 0.005 \\ 
 5 & J045747     &  14.226  & 15.386      &  14.817 & 0.569    & 13.121     & 12.186     & 10.810     &  8.573 &  7.657  &  6.936  &  5.799 & 2.193 \\ 
 & $\pm$       &   0.040  &  0.040      &   0.037  &   &  0.031     &  0.031     &  0.023     &  0.047 &  0.021  &  0.025  &  0.030 & 0.007 \\ 
6 & J045907      &  12.309  & 13.154      &  13.297 & $-$0.143    & 13.658     & 13.831     & 13.726     & 13.420 & 12.664  & 11.891  &  8.757 & 3.146 \\ 
& $\pm$       &   0.118  &  0.029      &   0.094  &   &  0.026     &  0.039     &  0.045     &  0.049 &  0.025  &  0.048  &  0.027 & 0.005 \\ 
7 & J051110     &  20.528  & 18.416      &  16.629 & 1.787    & 13.295     & 12.844     & 12.844     & 12.125 & 11.688  &  9.937  &  7.384 & 3.561 \\ 
 & $\pm$       &   0.192  &  0.050      &   0.033 &    &  0.028     &  0.031     &  0.030     &  0.043 &  0.045  &  0.036  &  0.027 & 0.007 \\       
 8 & J051228     &  15.598  & 16.586      &  16.154 & 0.432    & 15.682     & 14.927     & 14.254$^a$ & 13.420 & 12.664  & 10.121  &  8.239 & 3.646 \\ 
 & $\pm$       &   0.067  &  0.040      &   0.046  &   &  0.084     &  0.096     &  0.042     &  0.049 &  0.025  &  0.040  &  0.028 & 0.006 \\  
 9 & J052043    &  17.618  & 16.515      &  15.075 & 1.440    &            &            & 12.385     & 12.583 & 12.340  & 11.319  &  8.942 & 4.488 \\ 
 & $\pm$       &   0.070  &  0.056      &   0.086  &   &            &            &  0.045     &  0.059 &  0.059  &  0.037  &  0.025 & 0.009 \\  
 10 & J052229     &  16.327  & 16.291      &  16.576  & $-$0.285   & 14.255     & 13.642     & 13.292     &        & 12.221  & 11.067  &  9.263 & 4.042 \\ 
 & $\pm$       &   0.060  &  0.044      &   0.103 &    &  0.064     &  0.088     &  0.057     &        &  0.071  &  0.055  &  0.033 & 0.008 \\  
 11 & J052520     &  17.562  & 16.570      &  15.335 & 1.245    & 13.627     & 13.445     & 13.310     & 13.086 & 12.927  & 11.206  &  8.587 & 4.124 \\ 
 & $\pm$       &   0.040  &  0.030      &   0.020 &    &  0.032     &  0.035     &  0.039     &  0.041 &  0.034  &  0.056  &  0.031 & 0.009 \\  
 12 & J052915     &  15.945  & 17.093      &  16.628  & 0.465   & 16.620$^a$ & 16.521$^a$ & 15.881$^a$ & 14.454 & 13.908  & 12.369  & 10.726 & 5.655 \\ 
 & $\pm$       &   0.069  &  0.018      &   0.131  &   &  0.058     &  0.112     &  0.123     &  0.034 &  0.050  &  0.044  &  0.040 & 0.012 \\  
 13 & J053348     &          & 15.710$^b$  &  15.560$^b$ & 0.150 & 15.493     & 14.777     & 13.154     &  9.994 &  8.884  &  7.978  &  7.017 & 4.300 \\ 
 & $\pm$       &          &             &             &  & 0.082     &  0.096     &  0.046     &  0.030 &  0.042  &  0.030  &  0.024 & 0.007 \\  
 14 & J054055     &  13.685  & 14.466      &  14.375  & 0.091   & 14.276     & 14.152     & 14.122     & 14.001 & 13.430  & 12.923  & 10.093 & 4.428 \\ 
 & $\pm$       &   0.036  &  0.029      &   0.029  &   &  0.033     &  0.048     &  0.078     &  0.048 &  0.038  &  0.074  &  0.040 & 0.009 \\  
 15 & J055825     &  15.979  & 16.044      &  15.418 & 0.626    & 14.672     & 14.145     & 13.201     & 11.449 & 10.523  &  8.893  &  6.801 & 3.044 \\ 
 & $\pm$       &   0.048  &  0.024      &   0.022 &    &  0.034     &  0.047     &  0.041     &  0.040 &  0.029  &  0.033  &  0.022 & 0.008 \\       
 
 \hline
\end{tabular}
}
\begin{list}{}{}
\item UBV photometry for SMC objects (the first four entries) from \citet{2002AJ....123..855Z}; for LMC objects (except J053348) from \citet{2004AJ....128.1606Z}; for J, H, K from 2MASS or, in a few cases, from 2MASS 6X; for [3.6], [4.5], [5.8], [8.0], and [24] from SAGE LMC and SMC surveys. \\
 $^a$ 2MASS 6X catalogue \\
 $^b$ NOMAD catalogue \citep{Zacharias:2004aa}. 
\end{list}

\end{table*}

We made our observations with the Robert Stobie Spectrograph mounted on the South African Large Telescope (SALT).  Two settings were used. The red setting used the 900 l/mm grating and an 8 arcmin long, 1.2 arcsec wide slit. The spectrum covered the
wavelength range $6160-9140$\,\AA, with two gaps, $7130-7200$\,\AA\ and
$8170-8220$\,\AA, and a  resolution of R=1500 (1\,\AA/pix) at the
central wavelength. The blue setting used a  slit similar to that of  the red setting and
the 3000 l/mm grating.  It covered the wavelength range $3280-4090$\,\AA,
with the gaps $3550-3580$\,\AA\ and $3840-3850$\,\AA, and a resolution of
R=3000 (0.25\,\AA/pix) at the central wavelength. The log of observations is shown in Table~\ref{log}, which contains date of observations, the object running number (see 
 Table\,\ref{objlist} for these numbers),  short object name, setting, V magnitude, and the exposure time. Observations were repeated when the observing conditions did not meet the proposal criteria. Table\,\ref{photometry} contains photometry and errors of the observed sources at 
U, B, V, J, H, K, [3.6], [4.5], [5.8], 8.0], and [24] (see  the table notes for the source of the photometry).

\section{Results}

In the observed sample one object (J004747) has a rather featureless spectrum, ten objects show emission lines, or emission and absorption lines in their spectra (see Fig.\,\ref{allem}), while five objects show clear absorption-line spectra (including the standard star Sk 105) (see Fig.\,\ref{allabs}). We classified objects with the absorption lines
according to the MK system \citep{1984ApJS...56..257J} 
and performed quantitative analysis with atmospheric modelling to derive their basic physical parameters.

For emission-line
spectra, we analysed these line intensities and measured radial velocities of emission lines present
in the spectra in order to assign them to the relevant object class. We also
inspected the 2D spectra in order to look for diffuse emission that could result from proximity of star forming regions. Our results are summarised in 
Table \ref{objects}, which contains the object running number, short object name,  classification of the object according to the literature, spectral class obtained tentatively from the comparison of the spectra with the spectra by \citet{1995A&AS..112..475A}, 
the effective temperature determined from our atmosphere modelling, log of the surface gravity obtained from our atmosphere modelling,  and   remarks on the given object.

\begin{table*}
\caption{Summary of the results.}
\label{objects}
\centering
{\tiny
\begin{tabular}{lllllll}
\hline
\hline
No. & object            & literature class.     &our class.     &\Tef   &\logg  &remarks        \\ \hline
& Sk 105                &F0I [1]        & A7-F2Ib               &6500   &0.5    & \\
1 & J004747             & O7 [2]                    &   YSO     &           &        & no emission lines, very strong \de \\
2 & J004841         &YSO [2]    & YSO           &       &  &[S\,{\sc iii}],[N\,{\sc ii}],[S\,{\sc ii}],$\rm H\,\sc{i}$ emissions; \de \\
3 & J010546     &YSO [3]        & B8Ia/b; pAGB          & 10000 &3.0  &H\,{\sc i}; em object south to the target; no \de  \\
    &        &          &               &15000 &2.5   & \\
4 & J011542     &YSO [3]        &Herbig Ae/Be star?&        &       &H\,{\sc i},Fe\,{\sc ii} emissions;weak \de \\
5 & J045747         &pAGB cand [4]&Herbig Ae/Be star?   &       &           &[N\,{\sc ii}],H\,{\sc i} Fe\,{\sc ii} emissions; $\rm H\alpha$ with wide wings, \de \\
6 & J045907     &O9ep/B1-2ep [4]&O9e: massive star      &       &           &[N\,{\sc ii}],H\,{\sc i},[S\,{\sc ii}],[O\,{\sc ii}] emissions;He\,{\sc i} absorption;no \de \\
7 & J051110             &pAGB cand F3II(e) [4]&F0Ib     &6750   &0.5    &H\,{\sc i},Ba\,{\sc ii} and Li\,{\i}? absorptions; no \de \\
8 & J051228         & pAGB cand. [4] YSO cand. [5] &YSO &       &       &[N\,{\sc ii}],H\,{\sc i},He\,{\sc i} emissions; little \de \\
9 & J052043         &pAGB cand F5Ib(e) [4]&F5I  &6250   &0.5    & no \de \\
10 & J052229     &galaxy [4]    &galaxy &       &       & RVel of $\sim$10000 km/s \\
11 & J052520        &pAGB cand A1Ia [4]     &A2Ib&6750&0.5      &H\,{\sc i} absorption,strong Sr\,{\sc ii} 4077\AA\ line - I luminosity class?;\\
    &                       &                                       &        &        &       & no de\\
12 & J052915            &pAGB cand [4] PN [6]&PN        &       &       &[O\,{\sc i}],[S\,{\sc iii}],[N\,{\sc ii}],H\,{\sc i},[O\,{\sc ii}],[Ar\,{\sc iii}],[S\,{\sc iii}] em.; \\
    &                       &                                       &        &        &       & \de; em. object north to the target\\
13 & J053348            &R CrB type [7] WC [5]&R\,CrB   &       &       &C\,{\sc ii}, He\,{\sc i} em, no H$\alpha$?; no or little \de \\
14 & J054055    &O-B type star [4]&O9: massive star     &       &       & [S\,{\sc iii}] 9069.19\AA, He\,{\sc i} absorptions; weak \de \\
15 & J055825            &eAGB [8] WC [5]&R\,CrB &       &       &unidentified lines \& molecular bands in blue, \\
   &                        &                                       &        &        &       &similar to J053348 in red; no \de \\
\hline
\end{tabular}
}
\begin{list}{}{}
{\tiny{\item 
\de - background diffuse $\rm H \alpha$ emission is present in the field; 
[1] \citet{Neugent:2010aa}; 
[2] \citet{Sheets:2013aa};
[3] \citet{2007ApJ...655..212B};
[4] \citet{2011A&A...530A..90V}; 
[5] \citet{2009ApJS..184..172G}; 
[6] \citet{1978PASP...90..621S}; 
[7] \citet{1996ApJ...470..583A}; 
[8] \citet{2009AJ....137.3139V} }}
\end{list}

\end{table*}

\subsection{Emission-line objects}

In Fig.\ref{allem} we present the red spectra of the objects clearly showing emission, and in two cases (J045907, J054055) also showing absorption lines of He I. 
The objects are plotted in order of increasing R.A. There are three SMC and eight LMC objects in the plot. Below we discuss in detail their spectra.

\begin{figure*}
\resizebox{\hsize}{!}{\includegraphics[width=0.95\linewidth]{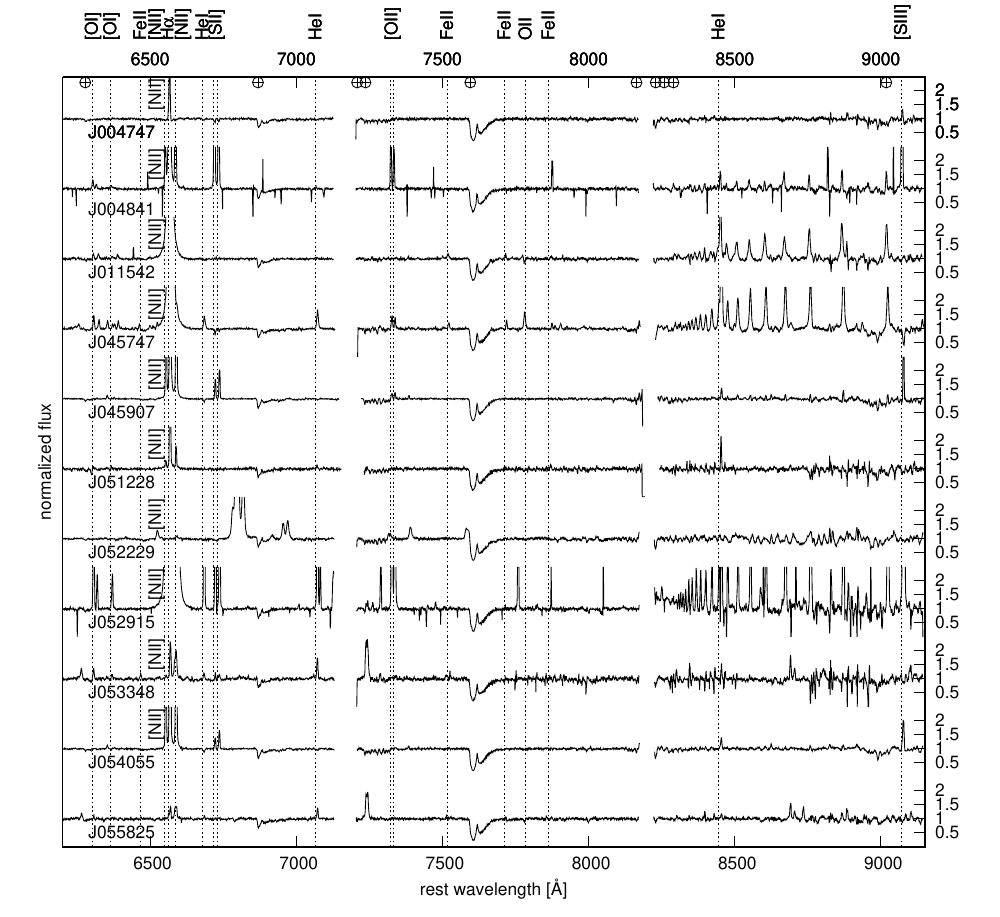}}
\caption{Red spectra of the objects showing emission lines or emission 
and absorption lines. The spectra were normalised to continuum. A constant offset of 2.5 was applied between the adjacent spectra. Telluric lines are indicated by   wheel cross symbols at the top of the plot. Vertical dashed lines give rest wavelength of the indicated emission lines. For clarity, lines representing Fe\,{\sc ii} emissions are not described.}
\label{allem}
\end{figure*}

\begin{figure*}
\resizebox{\hsize}{!}{\includegraphics[width=0.95\linewidth]{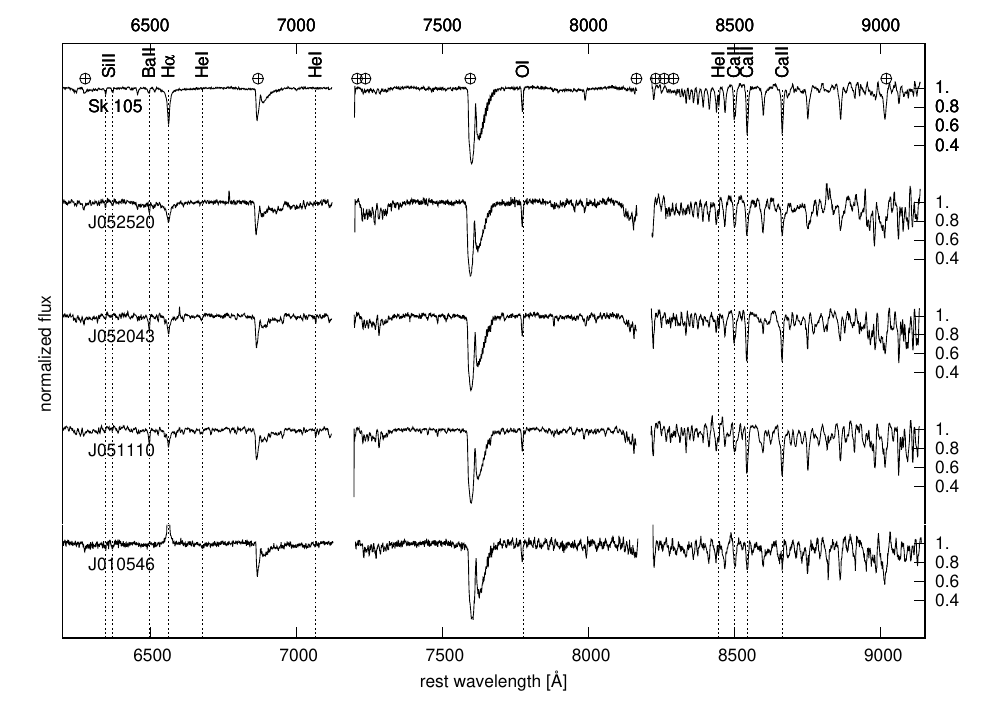}}
\caption{Spectra of the absorption line or featureless objects in the red setting. The spectra were normalised to continuum.  A constant offset of 1.2 was applied between the adjacent spectra. Vertical dashed lines indicate the rest wavelengths of the indicated absorption lines.}
\label{allabs}
\end{figure*}

{\bf Object No.1:} (see Table \ref{objlist}) {\bf J004747} 
is an O7 dusty star \citep{Sheets:2013aa}. The SALT spectrum  shows weak $\rm H \alpha$ emission superimposed on strong featureless
continuum (top spectrum in Figure \ref{allem}). 
Most probably it is residual emission not well subtracted from the background. No other spectral features have been identified in the spectrum allowing for the object classification. Diffuse background emission suggests that it is located in an emission region and is likely a relatively young object.

{\bf Object No.2: J004841} 
shows [S\,{\sc iii}] 6312\,\AA\ and 9069\,\AA\ lines and likely the [Ar\,{\sc iii}] 7751\,\AA\ line on a detectable continuum.
The object resides in a faint extended $\rm H \alpha$ emission region, and
was considered   a very low excitation (VLE) object by
\citet{1993A&AS..102..451M} and classified as a young stellar object (YSO) of B0 V? spectral type by \citet{Sheets:2013aa}. In addition, \cite{Oliveira:2013aa} classified this object as a candidate YSO. 

{\bf Objects No.4: J011542 
and No.5: J045747} have very similar spectra: 
both show very strong $\rm H \alpha$ emission, exceeding by far any other lines in the spectrum, with broad wings and no detectable [N\,{\sc ii}] 6583/6548\,\AA\ lines. Instead, they show Fe\,{\sc ii} and oxygen emission lines in the spectra. Paschen series lines in emission dominate in the red part of the spectra. These characteristics fits the criteria of the so-called Be phenomenon \citep{2000ASPC..214...26Z}.
However, these stars have strong IR excess, which is due to circumstellar dust not to free-free emission as in classical Be stars. Therefore, they are likely Herbig Ae/Be (HAeBe) stars. 
The diffuse background emission, which is detected in the spectra of both these objects (see Tab.\ref{objects}) and indicates that there is a lot of matter in the vicinity of both stars, supports their classification as young objects. 

J045747 has a Spitzer spectrum and \citet{Jones:2017aa} classified it as YSO4. As we have checked, its bolometric luminosity is of the order of 45\,000 solar luminosities, and there is no doubt that this is a massive Herbig Ae/Be star taking into account the above-discussed properties; we note, however, that van Aarle et al. 2011 
retains J045747 in the catalogue of post-AGB candidates.
On the other hand, \citet{2014MNRAS.tmp..343K} classified J011542 as a hot post-AGB/RGB candidate. We estimated that the bolometric luminosity is about 18\,000 solar luminosities, and in principle it could be a massive post-AGB object. However, the similarity of features observed in the optical spectra of both objects suggests that it is also a Herbig Ae/Be object.

{\bf Object No.6: J045907 
and No.14: J054055} 
both show strong $\rm H\,\alpha$ together with [N\,{\sc ii}] 6584/6548\,\AA\ emission, and 6678\,\AA\ and 7065\,\AA\ He\,{\sc i} absorption lines. In addition, emission lines of Si\,{\sc ii} 6347\,\AA\ and 6371\,\AA, [S\,{\sc iii}] 6312\,\AA\ and 9069\,\AA, [S\,{\sc ii}] 6716/6731\,\AA, and [O\,{\sc ii}] 7319/7330\,\AA\ 
are seen in the spectra. J045907 was classified as a O9ep/B1-2ep star by \citet{2011A&A...530A..90V} and was not considered a post-AGB star, due to its excessively
high luminosity. Our spectrum suggests an O type (possibly O9e:), but it relies
only on two He\,{\sc i} lines (6678\,\AA\ and 7065\,\AA) seen in absorption. Similarly, J054055
has been classified as O-B type by \citet{2011A&A...530A..90V}, and was not considered  a post-AGB star, due to its observed luminosity exceeding the upper limit expected for a post-AGB star. Our spectrum suggests  O type (possibly O9), but again it relies on only  two He\,{\sc i} absorption lines at 6678\,\AA\ and 7065\,\AA. 

{\bf Object No.8: J051228} 
shows only $\rm H\alpha$, [N\,{\sc ii}]
6584/6548\,\AA, He\,{\sc i} 7065\,\AA, and O\,{\sc i} 8446\,\AA\ lines superimposed on relatively strong continuum in its spectrum. A planetary nebula central star would have
a much weaker continuum. \citet{2011A&A...530A..90V} did not assign any spectral type for this object, but list it as a post-AGB candidate. However, a weak diffuse emission is detected in the vicinity of
the object, which suggests its young age. For this reason the object appears to be a YSO or H\,II region rather than a post-AGB star.

{\bf Object No.10: J052229} 
shows $\rm H \alpha$, [N\,{\sc ii}] 6584/6548\,\AA,
[S\,{\sc ii}] 6716/6731\,\AA,\ and other emission lines redshifted by $\rm {\sim}10000\,km\,s^{-1}$ with respect to the LMC (Figure \ref{allem}). We classify it as a background galaxy, in agreement with \citet{2011A&A...530A..90V}.

{\bf Object No.12: J052915} 
is listed as a post-AGB candidate
by \citet{2011A&A...530A..90V}, but it is a known PN \citep{Reid:2010aa}. Our observation supports this classification. The
object shows strong [N\,{\sc ii}] 6584/6548\,\AA, [O\,{\sc i}] 6300/6363\,\AA,
[S\,{\sc ii}] 6723/6737\,\AA, [O\,{\sc ii}] 7319/7330\,\AA, [S\,{\sc iii}]
9096/6312\,\AA,\ and [Ar\,{\sc iii}] 7751\,\AA\ lines in addition to the helium
and hydrogen lines. 

{\bf Objects No.13: J053348} 
{\bf and No.15: J055825} 
show very similar spectra (Figure \ref{allem}). \citet{1996ApJ...470..583A} classified J053348 as a hot R\,CrB star with a $\rm T_{eff}$ of about 20,000\,K. J055825 has been classified as an He-burning AGB star \citep{2009AJ....137.3139V}. Absorption lines or bands and
emission lines are observed in the blue part of the spectrum of
J055825 (Figure \ref{allblue}).
\citet{2011A&A...530A..90V}
classified both objects as Wolf--Rayet carbon stars and discard them from the
post-AGB sample. They noted, however, the resemblance of the spectra of both stars with
that of Galactic R\,CrB star V\,348 Sagittarius.

\begin{figure*}
\resizebox{\hsize}{!}{\includegraphics[width=0.95\linewidth]{{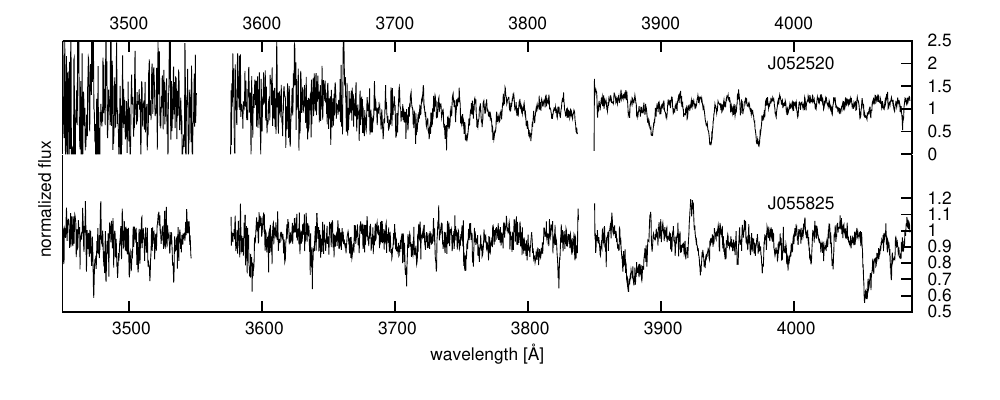}}}
\caption{Spectra of objects in the blue setting. The spectra are normalised to continuum.}
\label{allblue}
\end{figure*}

\subsection{Absorption-line objects}

In Fig.\ref{allabs} we present the red spectra of the five objects with absorption lines in their spectra. First the spectrum of our standard star, Sk 105, is plotted, and then objects No. 11, 9, 7, and 3 (the only object from the
SMC). We put them in order of decreasing R.A.  to facilitate further discussion. We present our MK classification of the spectra and by comparing the synthesised colours for a given spectral class with observed values we derive total (circumstellar plus interstellar) extinction. Then we present quantitative analysis with the model atmosphere and spectra synthesis technique to derive basic physical parameters of the objects.

\subsubsection{MK classification}
\label{_MK}

We performed the MK classification according to \citet{2009ssc..book.....G} and standards observed by \citet{1995A&AS..112..475A}. Our MK classifications relies mainly on the strength of the Ca\,{\sc ii} triplet 8498\AA, 8542\AA, and 8662\AA, blended with P16, P15, and P13 Paschen lines, relative to other lines from the Paschen sequence, which is sensitive to stellar temperature \citep{2009ssc..book.....G}. In supergiants the Paschen lines are narrower with respect to other luminosity classes. However, it should be noted that the standards observed by \citet{1995A&AS..112..475A} are mainly (or exclusively) population I stars with higher metallicities than our targets, thus  more relevant for comparison with Galactic objects. The decreased metallicity in MC objects results in reduced strength of the Ca\,{\sc ii} lines. Thus, the spectral types and stellar temperatures derived from MK classification based on the relative strength of the Ca\,{\sc ii} lines would be systematically higher. This poses the main problem and main source of uncertainties in our classification.
The spectral resolution of our observations in the red setting and observations by \citet{1995A&AS..112..475A} is quite similar (about 1\,\AA) and should not introduce additional uncertainties. 

All  five objects show hydrogen Paschen series in absorption. The lines are relatively narrow and deep, which results in lines from upper levels, including those above P17, to be visible. This indicates a supergiant luminosity class. The lines identified in the spectra are listed in Tab. \ref{objects}. The detailed discussion of the spectral type classification of each is given below.

{\bf Sk\,105} is a yellow supergiant of MK spectral type F0\,I \citet{Neugent:2010aa}.
Our spectrum allowed us to constrain the spectral type between A7 and F2, and luminosity class Ib. The photometric B-V colour index (see Tab.3) match  this spectral type quite closely, assuming negligible circumstellar or interstellar extinction, and the corresponding effective temperature for the assigned F0Ib spectral type is 7460\,K \citep{cox2000}.

{\bf Object No.11: J052520} 
is a C-rich post-AGB star with a strong $21\,\mu$m dust feature \citep{2011ApJ...735..127V, Matsuura:2014ve, Sloan:2014uq}. This object varies with a periodicity between 34 and 45 days, but does not have a stable dominant period \citep{Hrivnak:2015aa}. From their low-resolution spectra, \citet{2011A&A...530A..90V} determined the spectral type of this object as A1Ia \citep[about 9700\,K; ][]{cox2000}, and estimated its effective temperature from the SED to be 9250 $\pm$ 250 K. However, \citet{Matsuura:2014ve} derived a spectral type of F2-F5\,I, with reference to the spectral atlas of \citet{1984ApJS...56..257J}, based on strong Ca\,II H and K, weak G band, sharp weak H line absorption, and strong O\,I 7774\,\AA\, seen in their spectrum.

J052520 was observed in the red and blue settings  with SALT (see  blue spectrum   in   top panel of Fig.\,\ref{allblue}). We classify this star as an A2Ib type \citep[about 9000\,K; ][]{cox2000}. Clearly, the Ca\,{\sc ii} triplet is less pronounced than in Sk\,105, indicating earlier spectral type. Strong Ca\,II H and K lines present in the blue part of the spectrum indicate, on the other hand, A0 type or later. The spectrum shows strong Sr\,{\sc ii} 4077\,\AA\ line, confirming luminosity class I. J052520  clearly shows an s-process Ba element line at 6496\,\AA\ (see further discussion). The observed colour B-V=1.25 \citep{2004AJ....128.1606Z}, with a spectral type of A2I \citep[(B-V)$_0$=0.03;][]{cox2000} indicates a total (circumstellar plus interstellar) reddening E$_{B-V}$ of about 1.2. 

{\bf Object No.9: J052043} 
is another C-rich post-AGB star with a strong $21\,\mu$m dust feature \citep{2011ApJ...735..127V, Matsuura:2014ve, Sloan:2014uq}. It has a complicated variability pattern with dominant period of about 74 days \citep{Hrivnak:2015aa}. This object is classified as a F5Ib(e) type \citep[about 6400\,K;][]{cox2000} by \citet{2011A&A...530A..90V}, or F8I(e)  \citep[about 5750\,K;][]{cox2000} by \citet{Hrivnak:2015aa}. J052043  clearly shows an s-process Ba element line at 6496\,\AA\ (see further discussion). We assign F5I spectral type \citep[about 6370\,K;][]{cox2000}
for this object based on the strength of Ca\,{\sc ii} triplet compared to Pa series. 
The observed colour B-V=1.44 \citep{2004AJ....128.1606Z}, with a spectral type of F5I \citep[(B-V)$_0$=0.32;][]{cox2000} indicates a total reddening E$_{B-V}$ of about 1.1.

{\bf Object No.7: J051110} 
is also a C-rich post-AGB star with $21\,\mu$m dust feature \citep{2011ApJ...735..127V, Matsuura:2014ve, Sloan:2014uq}. This object shows clear evidence of variability with two significant periods of about 112 and 96 days \citep{Hrivnak:2015aa}. \citet{2011A&A...530A..90V} determined the spectral type of this object as   F3II(e) type \citep[about 6800\,K for luminosity class I;][]{cox2000}, or slightly more for luminosity class II, which agrees with \Tef of 7000 $\pm$ 250 K obtained from their SED modelling. We classify it as F0Ib type \citep[about 7460\,K;][]{cox2000},
but with large uncertainty.
J051110 also shows  an s-process Ba element line at 6496\,\AA\ (see further discussion). The observed colour B-V of about 1.8 \citep{2004AJ....128.1606Z}, with a spectral type of FOIb \citep[(B-V)$_0$=0.17 for luminosity class I object;][]{cox2000} indicates a total reddening E$_{B-V}$ of about 1.6. Clearly, the object is heavily reddened.

{\bf Object No. 3: J010546} 
is the only SMC object in our sample with stellar absorption lines in its spectrum. It shows relatively weak $\rm H \alpha$ emission with broad wings superimposed on continuum, and some weak absorption lines. The Ca triplet does not appear to be superimposed 
on a Paschen series. The O\,I line at 7774\,\AA\ is visible, as are the Si II 6347, 6371 \,\AA\ lines. The best fit to the standards was achieved for B8 Ia/b type (temperature of about 11,100\,K;  \citealt{cox2000}). The observed colour indicates the total reddening E$_{B-V}$ of about 0.4 for this object. This source also  shows   the $21\,\mu m$ feature \citep{2011ApJ...735..127V, Sloan:2014uq}, so it is a C-rich post-AGB star. \citet{2014MNRAS.tmp..343K} list it as a hot post-AGB/RGB candidate. 

\subsubsection{Synthetic spectra: analysis procedure}

A grid of model atmospheres of supergiants\footnote{The model atmospheres are available from //ftp.mao.kiev.ua/pub/y/2017/postAGB/.} was computed using the SAM12  program
\citep{2003ARep...47...59P}. SAM12 is a modification of the  ATLAS12 code of
\citet{1993KurCD..13.....K}.  When computing cooler model atmospheres (\Tef$\leq$9000\,K), we modelled the
convective energy transport using mixing-length theory with the ratio of mixing length to scale height $l/h = 1.6$. The
line opacity was used in the framework of opacity sampling technique
\citep{1976ApJ...204..281S} with the VALD2 atomic list \citep{1999A&AS..138..119K}. 

We computed model atmospheres of effective
temperatures \Tef$=5250-8000$ K with steps of 250\,K and \Tef$=9000$ K; gravities 
\logg$=0, 0.5, 1$ for the cooler models and \logg$=1.5-2.5$ for the hotter models (step of 0.5 in \logg); and metallicities [Fe/H]$ = 0, -0.2, -0.5, -1, -1.5, -2, -2.5$.
Since these objects are all C-rich, we then computed model atmosphere grids for  
a set of [C/O] = +0.2, +0.4. We found that the dependence of our results 
on [C/O] is  marginal; nevertheless, we carried out our analysis for different values of [C/O].
For analysis of the hottest star in our sample we computed an additional set 
of hot model atmospheres of \Tef from 10\,000 to 20\,000\,K, with step of 1000\,K and 
\logg$= 2.0-3.5$.

The technique of synthetic spectra was used to carry out our  analysis of the
observed spectra computed  with the WITA6  program
\citep{1997Ap&SS.253...43P} within the classical framework. A microturbulent velocity of 
\Vt =3.5 km/s was adopted.
Our synthetic spectra were computed with wavelength steps
of 0.025 \AA, which provided detailed profiles of all the lines. 

In the most general case the quality of the fits to the observed spectra
depends on many adopted parameters: effective temperature, 
gravity,
microturbulent velocity,
and the abundances of elements. 
In this paper we solve a restricted case using fits to the selected narrow spectral
region 8450$-$8800\,\AA.
Unfortunately, we do not have good enough data to provide the detailed numerical analysis of the abundance of all elements. Therefore, we used the  parameter of metallicity for most of the elements, except for carbon and s-elements. Some s-process elements are so overabundant that their lines are seen in the spectra of the absorption-line  objects. 
This allowed us to estimate their abundances (see Section \ref{sab}).

The observed spectra are broadened by macroturbulence and by
instrumental broadening; both are implemented in the modelling by smoothing the
computed spectra with a correspondent Gaussian, which is a reasonable approach
for fitting stellar spectra.  The appropriate values of full width at half maximum were
determined from every spectrum during the procedure of finding the best fit to the observed spectrum.

Using the standard $\chi^2$ procedure to determine the best fit of synthetic spectra (S$_{\rm synt}$) from our grid
to the observed spectrum (S$_{\rm obs}$), we find the minimum of
the 3D functional
$$S(f_{\rm h} , f_{\rm s} , f_{\rm g}) = \Sigma(S_{\rm synt} (f_{\rm s}, f_{\rm g})\times f_{\rm h} - S_{\rm obs})^2,$$ 
where $f_{\rm h}$ is the flux normalisation coefficient, $f_{\rm s}$ is the
wavelength shift between the theoretical and observed spectra, and 
$f_{\rm g}$ a parameter of convolution describing the broadening by macroturbulent motions and instrumental
broadening. The observed spectra had been normalised to the continuum/pseudo-continuum prior modelling.

\subsubsection{Fits to the observed spectra}
\label{_fits}

The parameters of the stellar atmospheres were obtained
from the fits to the H\,I Paschen lines and the Ca\,{\sc ii} triplet at 8498\,\AA,
8542\,\AA,\ and 8662\,\AA\ lines. All three Ca lines were blended with the 
H\,{\sc i} Paschen lines at the resolution of our red spectra. This is demonstrated in Fig.\,\ref{Hlines}, which shows an example of a synthetic spectrum without hydrogen (red line) where the Ca\,{\sc ii} triplet lines are clearly seen, and a spectrum of pure hydrogen (green line) where the subsequent lines of the H Paschen series are identified.    
It is worth noting that for a given metallicity, Ca lines become weaker on the hot edge of our temperature range, while the strength of H lines
increases at the same time.
Thus, this spectral region provides a good opportunity for the self-consistent
determination of \Tef and [Fe/H]. For simplicity, we made the reasonable 
assumption that the abundance of Ca follows the trend of [Fe/H]. 

\begin{figure}
\resizebox{\hsize}{!}{\includegraphics{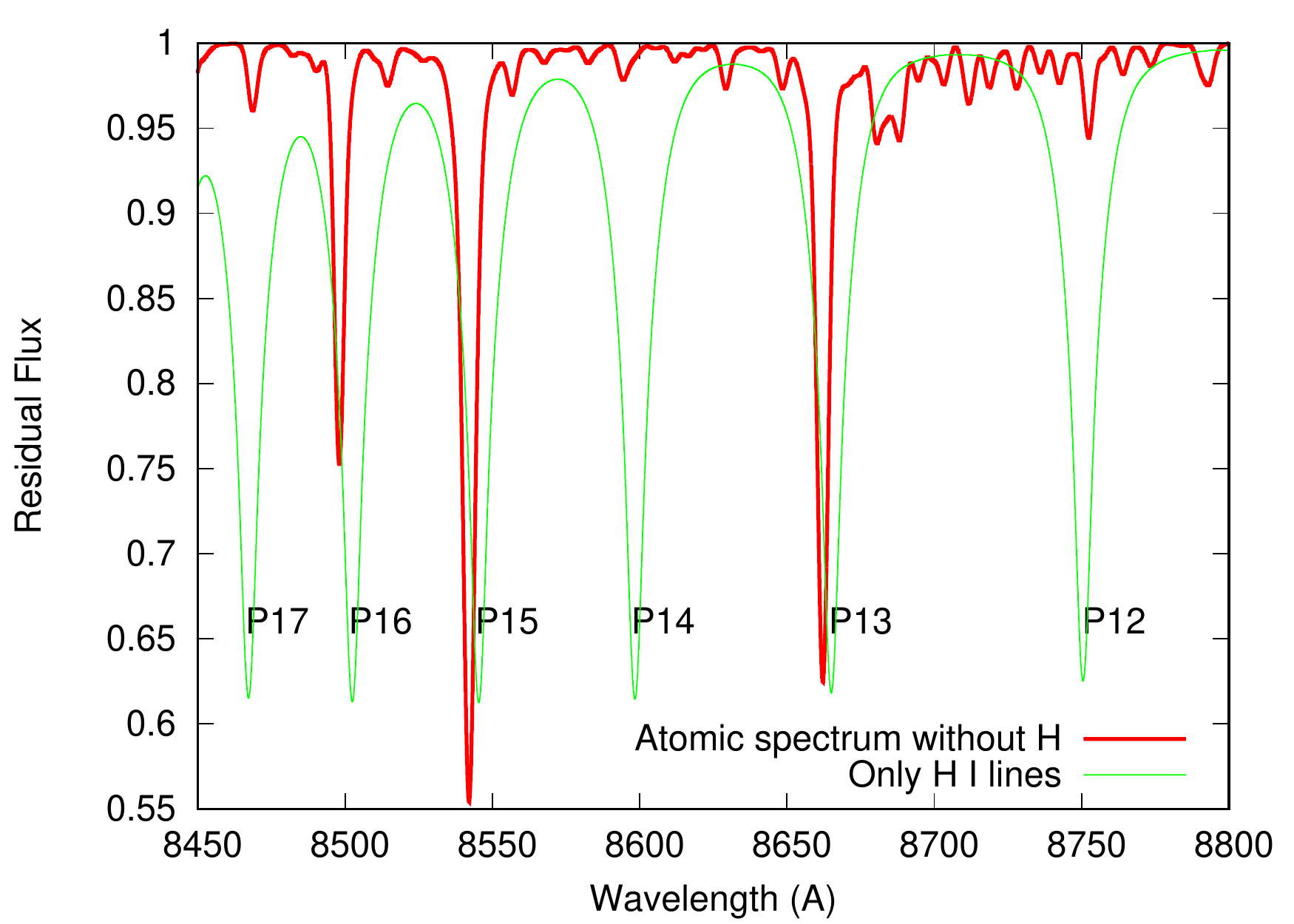}}
\caption{Synthetic spectrum computed from a model atmosphere without hydrogen (red line), and synthetic spectrum of pure hydrogen atmosphere (green line).  The Ca\,{\sc ii} triplet 8498\,\AA,
8542\,\AA,\ and 8662\,\AA\ lines are clearly seen in the red spectrum. The subsequent transitions of the hydrogen Paschen series seen in the green spectrum are indicated in the plot.}  
\label{Hlines}
\end{figure}

All observed spectra were affected by fringing, which possessed an amplitude of 
about 10\% on the red side. This restricted the accuracy of our fit. 
Nevertheless, using the procedure of fitting the observed spectra in the
spectral region of the Ca\,{\sc ii} and Paschen lines, we were able to determine \Tef and $\log g$.  
In our case, the formal error in the
measured  quantities should be equivalent to the lowest step in the grid, 
namely 250 K in \Tef and 0.5 in $\log g$.

Due to the broad profiles and blending effects of the  Paschen H\,{\sc i} 
photospheric  lines, it is very difficult to perform proper normalisation of the
spectra with respect to the local continuum level for the Paschen series  lines.
Thus, after preliminary modelling we  normalised the spectra  to the
continuum for a second time. Then the modelling was performed  again to get
more accurate stellar parameters.

We began by testing our modelling procedure by fitting the spectrum of the supergiant  Sk\,105, observed by us as a template star. Sk\,105 is
the brightest star in our sample and was observed with the best S/N.
Results of the fit to the Sk\,105 spectrum are
shown in Fig \ref{sk105}. The best fit to the observed spectrum, determined primarily by fitting the H
Paschen lines and Ca II subordinate triplet at 8498, 8542, and  8662
\AA, provides our estimation of its effective temperature, gravity,
metallicity, and [C/O]: 6500/0.5/-0.5/0.0.
Our estimation of \Tef is quite similar to the value of \Tef = 6800\,K 
derived by \citet{Neugent:2010aa}.
The presence of Paschen 
lines of H I and lines of the Ca II triplet of the observed strength qualitatively agree with our \Tef estimation.

\begin{figure}
\resizebox{\hsize}{!}{\includegraphics{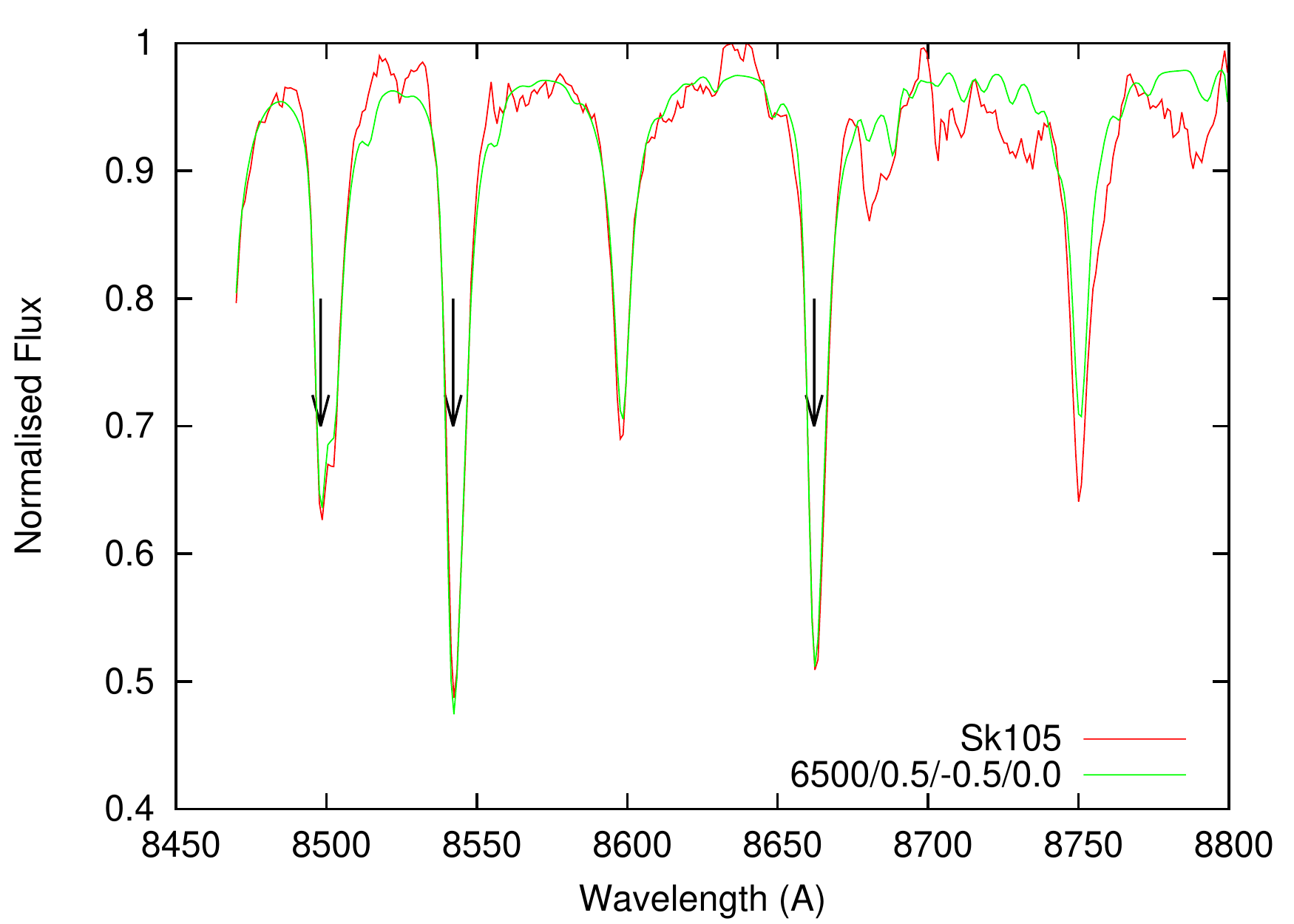}}
\caption{Best fit of the synthetic spectrum computed for a model atmosphere with effective temperature, gravity, metallicity, and [C/O]:
6500/0.5/-0.5/0.0 (green line) to the observed spectrum of Sk 105 (red line). Arrows indicate the Ca\,{\sc II} lines, which are  blended with H\,{\sc I} Paschen lines.}
\label{sk105}
\end{figure}

Following the reasonably good fit to the spectrum of Sk\,105, we applied our procedure to the four objects with absorption lines in their SALT spectra. Our best fits, together with the obtained model atmosphere parameters, are shown in  
Fig. \ref{J052520} for {\bf object No.11: J052520}, in Fig. \ref{J052043} for  {\bf object No.9: J052043}, in Fig. \ref{J051110} for {\bf object No.7: J051110},  and in Fig.\ref{J010546} for {\bf object No.3:  J010546}. The obtained model atmosphere parameters confirm the post-AGB nature of each object, and they are the first such values determined for J051110 and J010546. These model atmosphere values are listed in Table\,\ref{modatmpar}. 

\begin{table}
\caption{Parameters of the model atmospheres} 
\label{modatmpar}
\centering
\begin{tabular}{lllrl}
\hline
\hline
object   & T$_{\rm eff}$                  &  \logg & [Fe/H] & [C/O] \\ 
         & [K]               & dex     & dex    & dex \\ \hline
J052520   & 6750              & 0.5     & -1.5   & 0.4 \\
J052043 & 6250           &  0.5     & -0.2   & 0.2 \\
J051110 & 6750           &  0.5     & -1.0   & 0.4 \\
J010546 & 10000          &  3.0     & 0.0   & 0.2 \\
        & 15000          &  2.5     & 0.0   & 0.4 \\
\hline
\end{tabular}
\end{table}

\begin{figure}
\resizebox{\hsize}{!}{\includegraphics{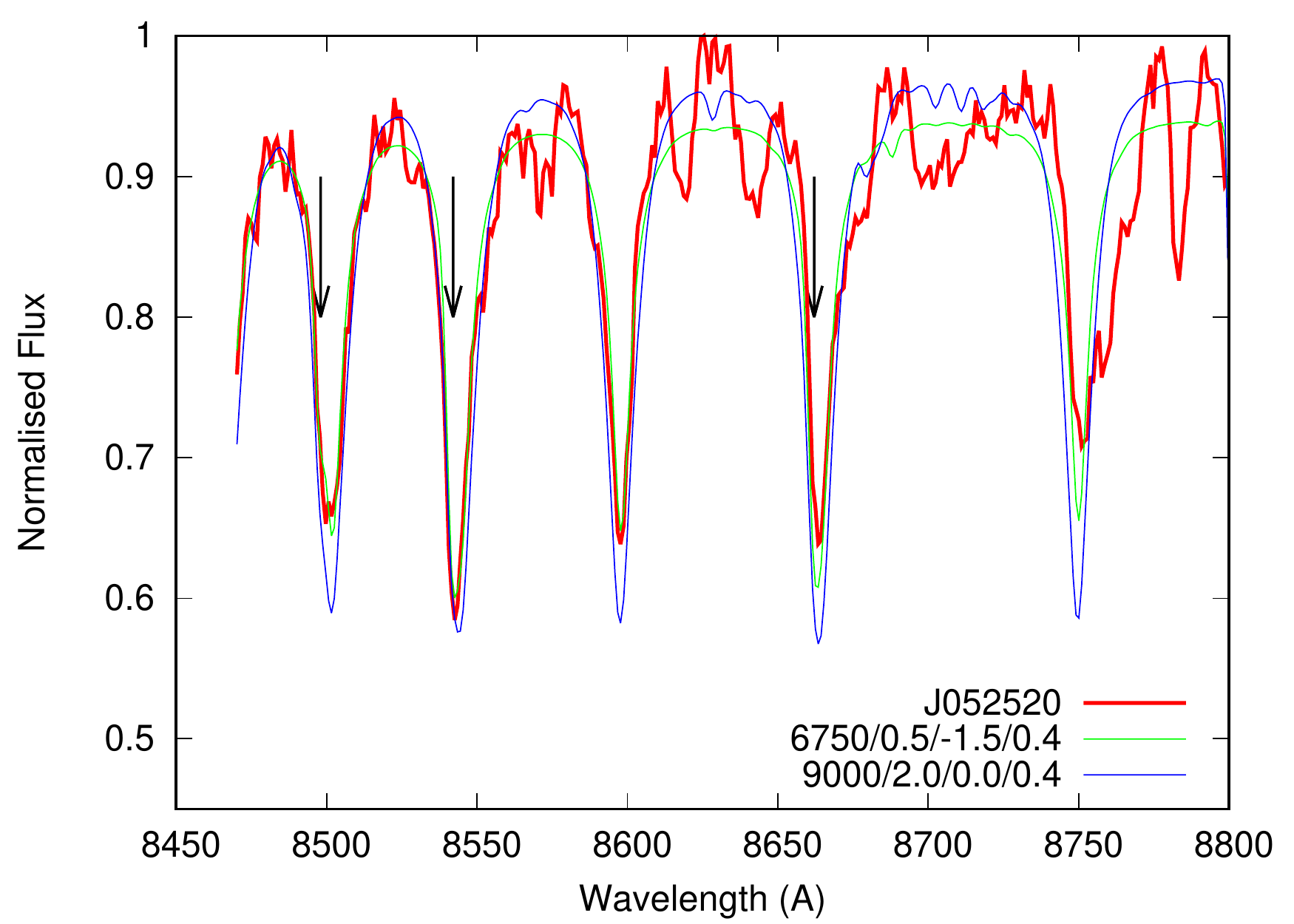}}
\caption{Best fit of the synthetic spectrum computed for a model atmosphere of
6750/0.5/-1.5/0.4 (green line) to the observed spectrum (red line) of object No.11: J052520. The blue line shows the best solution for the theoretical spectrum computed for model atmospheres with \Tef=9000.}
\label{J052520}
\end{figure}
\begin{figure}
\resizebox{\hsize}{!}{\includegraphics{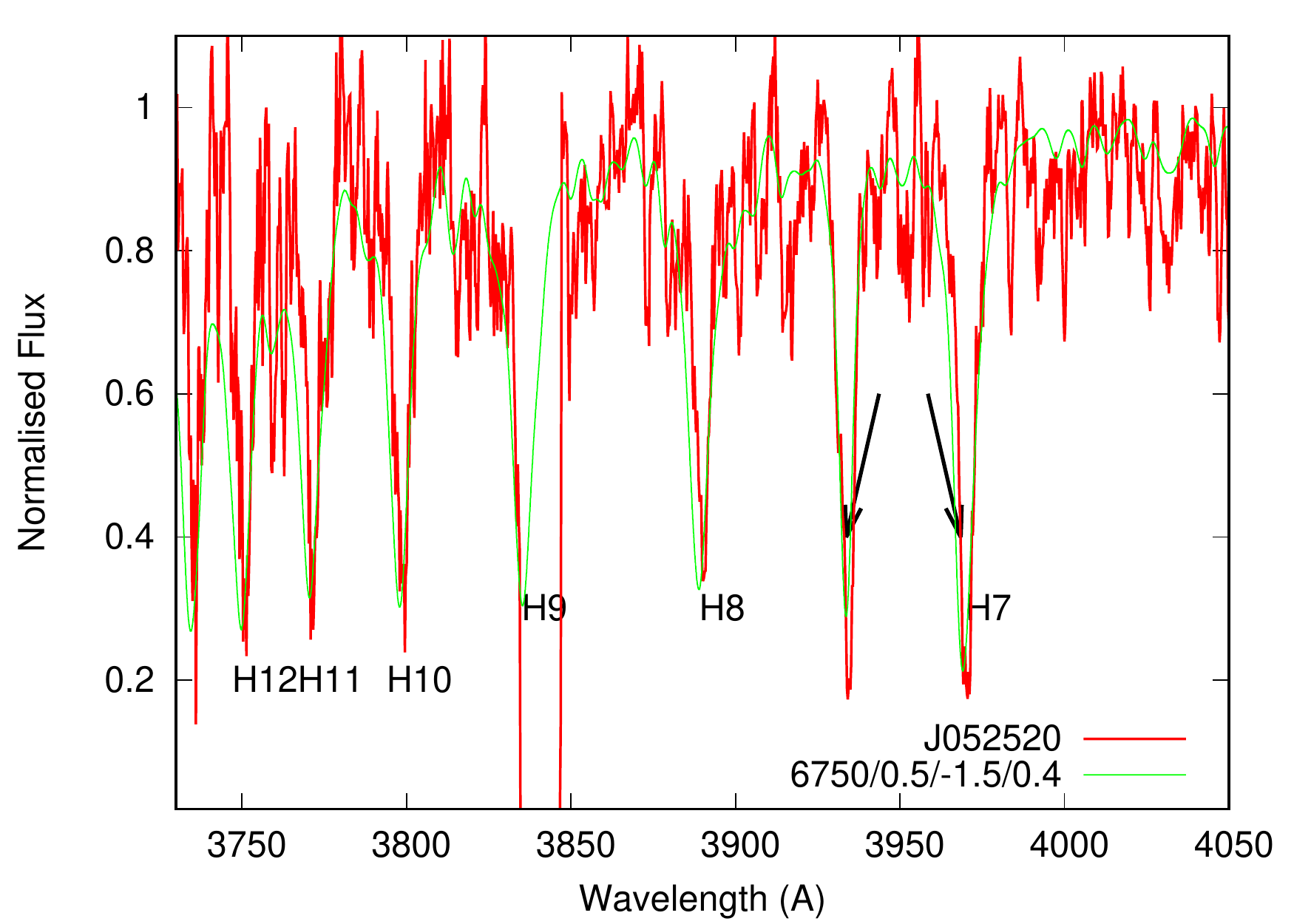}}
\caption{Fit to the blue part of spectrum (red line) of object No.11:  J052520 with stellar model parameters derived from the best fit to the red spectrum of this object. The Ca II resonance lines are indicated by arrows.}
\label{J052520blue}
\end{figure}
\begin{figure}
\resizebox{\hsize}{!}{\includegraphics{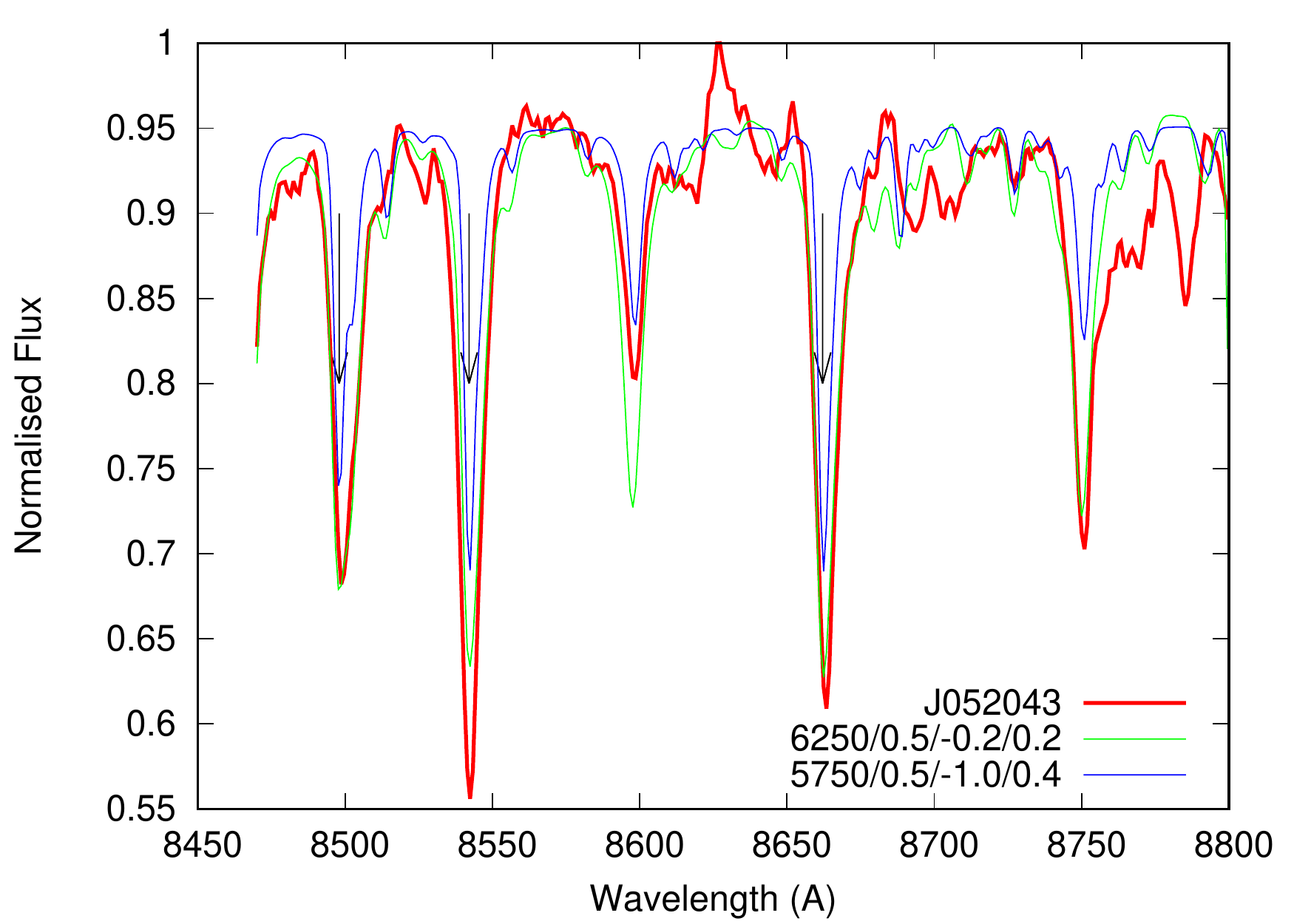}}
\caption{Best fit of the synthetic spectrum computed for a model atmosphere of 
6250/0.5/-0.2/0.2 (green line) to the observed spectrum (red line) of object No.9: J052043.
The blue line shows the fit to the observed spectrum with parameters 5750/0.5/-1.0 found by \citet{van-Aarle:2013fj}.}
\label{J052043}
\end{figure}
\begin{figure}
\resizebox{\hsize}{!}{\includegraphics{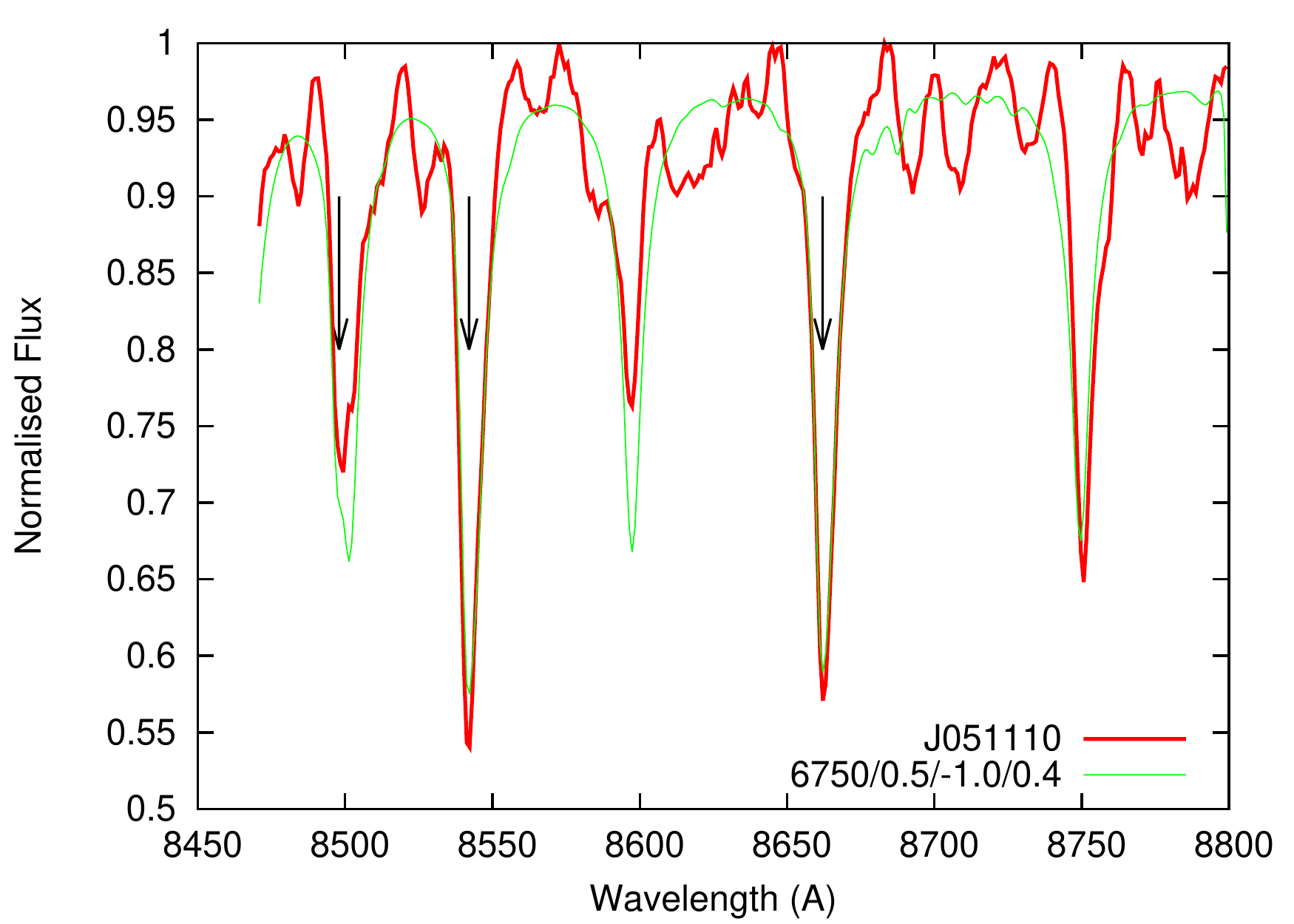}}
\caption{Best fit of the synthetic spectrum computed for a model atmosphere of
6750/0.5/-1.0/0.4 (green line) to
the observed spectrum of object No.7: J051110.}
\label{J051110}
\end{figure}

For {\bf J052520} the fit to the observed spectrum is fairly good (see Fig.\ref{J052520}) and the correctness of the parameters is supported by the good agreement of the fit to the blue part of the spectrum (see Fig.\ref{J052520blue}).
The blue part of the spectrum shows strong resonance lines of Ca II at
$\lambda\lambda$ 3934 and 3968\,\AA, which can be used as an independent check of the obtained best stellar parameters from the fits to lines in the near-IR spectrum.
In Fig. \ref{J052520blue} we compare observed blue part of the spectrum (red line) with the theoretical spectrum (green line) obtained for the best fitting parameters (see Table\,\ref{modatmpar}). As can be  seen, the theoretical spectrum reproduces fairly well the observed Ca II lines, and the H Paschen lines, thus providing evidence that our method gives consistent
results and that the obtained stellar parameters are correct.

\citet{Kamath:2015aa} has also derived stellar parameters from model atmosphere fitting to their low-resolution spectrum of this object. While their effective temperature is the same as in our model (6750\,K), the metallicity and especially the gravity differ. They  obtained \logg = 2.5, causing them to classify the star as a young object. Since the post-AGB nature of this object is well established (presence of 21\,$\mu$m feature), they argue that a large uncertainty in their \logg value arises from the difficulty in fitting the Ca triplet region in spectra of this temperature and the rather noisy spectrum in the region of the Balmer lines 3750 - 3950 \AA. 

The obtained \Tef of 6750\,K is very different to that inferred from the MK classification of about 9000\,K (see Sect.\,\ref{_MK}). In Fig.\,\ref{J052520} we also show the best fit obtained with 9000\,K. Clearly this temperature value results in hydrogen lines that are too strong. Taking into account all the above arguments we can conclude that the MK classification gives the wrong spectral type. Modelling yields the lowest metallicity for J052520 of all observed absorption-line objects, and this may be the reason for the large difference between MK classification and the model.

In the case of {\bf J052043}, the fit to the spectrum is good except for the Paschen line at ~8600 \AA~ (see Fig.\ref{J052043}).  Our spectral classification of F5I (about 6370\,K) is consistent with the derived temperature of 6250 K.
\citet{van-Aarle:2013fj}   derived a lower temperature (5750\,K) and lower metallicity ([Fe/H]=-1.0) from the analysis of their UVES spectra. In  Fig.\,\ref{J052043} we show  the synthetic spectrum computed for  parameters obtained by \citet{van-Aarle:2013fj}. As can be  seen,  the hydrogen lines are too weak for \Tef = 5750 K, and the the Ca lines are shallower for a metallicity [Fe/H] = -1.0, as expected. One explanation for this difference could be variability of the source. J052043 has a dominant period of about 74 days \citep{Hrivnak:2015aa}. 
\cite{Hrivnak:2010aa} found that PPNe show colour changes corresponding to temperature changes of 300$-$700\,K.

The fit for {\bf J051110} is reasonably good (see Fig.\,\ref{J051110}), but it does not fit  the Ca 8500\,\AA\ line or the Paschen line at 8600\,\AA.  The fit gives a temperature of 6750\,K.  This disagrees  quite a bit with the temperature estimation of ~7460\,K based on our MK classification as F0I. However, as we   mention  in Sect.\,\ref{_MK}, this determination was made with a large uncertainty. On the other hand, the difference can be accounted for by stellar pulsations and related variability in temperature. We note that
 the value of\Tef determined by our model agrees within the error with the temperature determination from MK classification or SED modelling by \citet{2011A&A...530A..90V}.

\begin{figure}
\resizebox{\hsize}{!}{\includegraphics{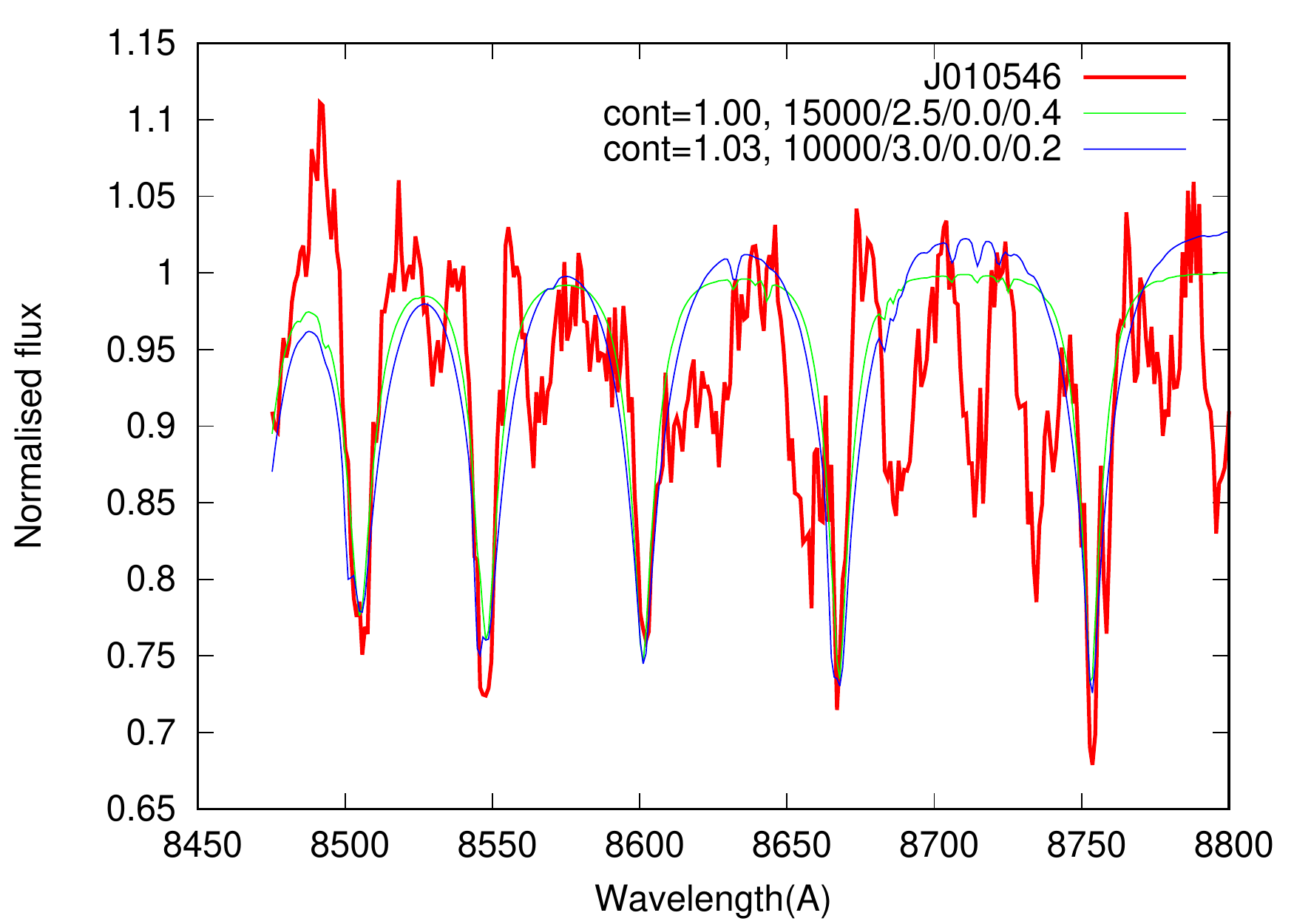}}
\caption{Best fit of the synthetic spectra computed for model atmospheres of 
15000/2.5/0.0/0.4 (green line) and 10000/3.0/0.0/0.2  (blue line) 
to the observed spectrum (red line) of object No.3 J010546, assuming
different continuum levels at 1.0 and 1.03, respectively.}
\label{J010546}
\end{figure}
\begin{figure}
\resizebox{\hsize}{!}{\includegraphics{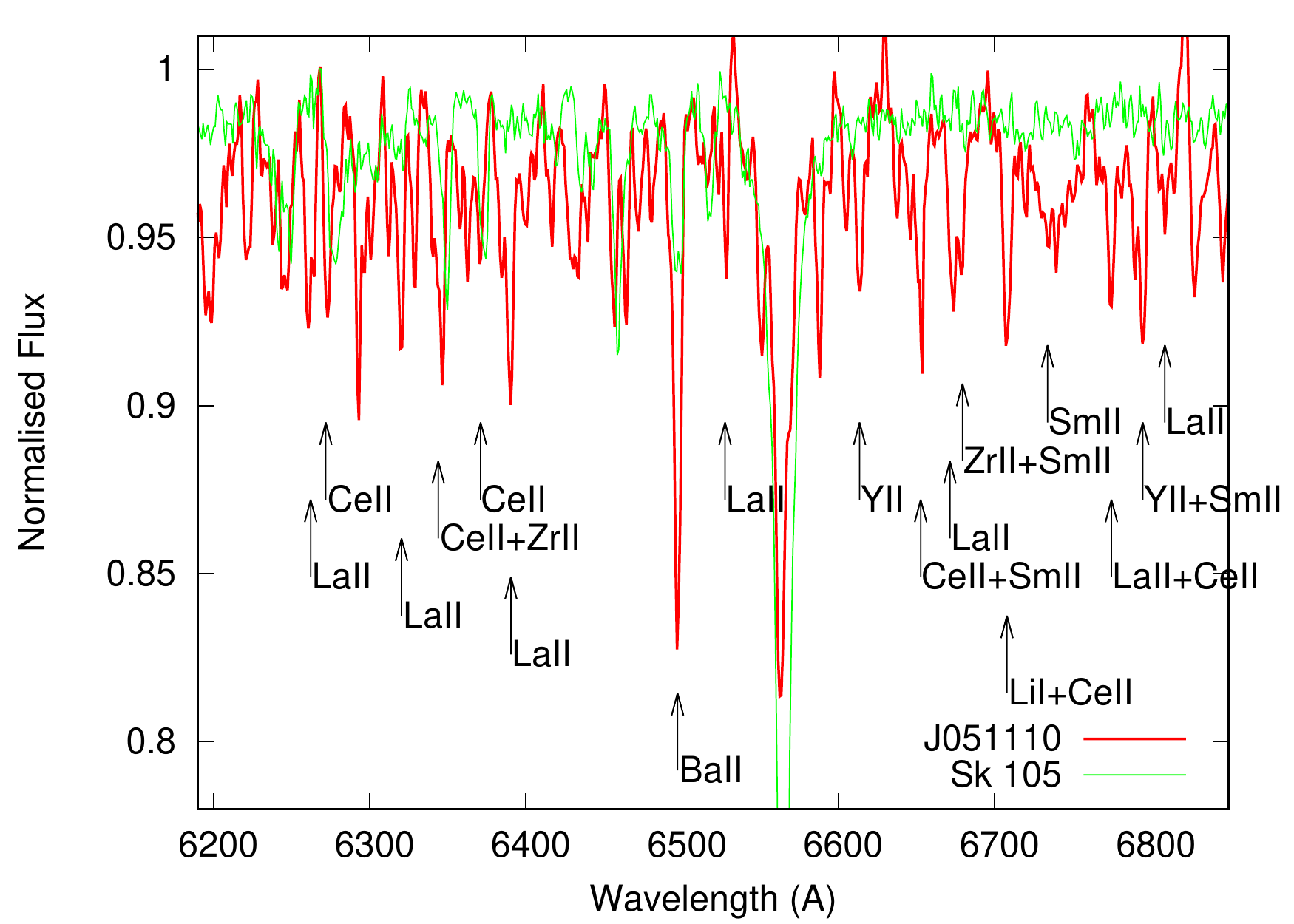}}
\caption{Comparison between the spectra of J051110 (red line) and the standard star Sk 105 (green line). Positions of some s-process element transitions are identified on the plot.}
\label{sel}
\end{figure}

{\bf J010546} is the hottest star among our absorption-line objects. It shows relatively weak
lines of hydrogen and ionised calcium, which both indicate a rather high effective 
temperature. Our fit indicates that the temperature of J010546 is in the range 10\,000$-$15\,000 K
 (see Fig. \ref{J010546}), which agrees with the MK classification of B8I (11100\,K). In general, the results of the best fit depend on the adopted continuum
level in the selected spectral range. For this source the structure of the spectrum is quite complicated, and the continuum level is poorly determined. Therefore, we performed fits for two
adopted continuum levels in the observed spectrum. Results for the continuum at 1.0 and 1.03 are shown in Fig. \ref{J010546} as the green and blue line, respectively.
Both models reproduce the P12 - P15 lines of hydrogen and the Ca II lines fairly well. However, the formal goodness $S=0.750 \pm 0.003$ is slightly better for the fit with \Tef=15000\,K than $S=0.890 \pm 0.004$ for \Tef of 10000\,K.

\subsubsection{s-elements abundance.}
\label{sab}
\begin{table}
\caption{Abundances of s-elements in the atmospheres of three post-AGB stars.} 
\label{_ab}
\begin{tabular}{cccccc}
\hline
\hline
       & Z & J051110 & J052043 &  \multicolumn{2}{c}{J052520} \\ \relax
       &  &   &   &     blue   &   red \\
\hline
[Y/Fe] & 39 & +2.2 &  +2.5 &    +2.5   &    +2.7  \\ \relax
[Zr/Fe]& 40 & +2.5 &  +2.5 &    +2.0   &    +2.0: \\ \relax
[Ba/Fe]& 56 & +3.3 &  +2.5 &    --     &    +2.9  \\ \relax
[La/Fe]& 57 & +3.0 &  +2.5 &    +2.0   &    +3.0  \\ \relax
[Ce/Fe]& 58 & +3.0 &  +1.8 &    +1.5   &    +2.0: \\ \relax
[Sm/Fe]& 62 & +2.6 &  +1.8 &    +1.5   &    +1.9: \\ 
\hline
\end{tabular}
\begin{list}{}{}
\item The `:' flag indicates   less confident results.
\end{list}

\end{table}
As we already noted, the resolution of our spectra is not high enough to carry out 
accurate 
determination of abundances. However, we see evidence of a significant 
contribution of s-element lines in the spectra of our cooler absorption-line stars. As an example, in Fig.\ref{sel}, we compare the 
spectrum of J051110, 
which is shown by a red line, with the spectrum of our standard star Sk 105 (green line) in the range from 6200 to 6850\,\AA.
The positions of some transitions of s-process elements (Ba, Ce, La, Sm, Y, and Zr) are shown by vertical arrows. As can be  seen, the spectrum of J051110 clearly shows absorption lines of s-process elements. 
 
To get reliable fits to the observed spectra of our stars in the wavelength range where s-process elements are clearly seen, we increased the abundance of s-process elements during computations of the model atmospheres (see
Table \ref{_ab}). Computations were carried out for the model atmospheres with parameters found from the fits to hydrogen
Paschen and Ca II lines (Table\,\ref{modatmpar}). However, while fitting the atomic lines of intermediate strength in the observed spectra, we found that a microturbulent velocity $V_t$=4.5\,km/s better describes the observations of s-element absorption lines.  Our numerical experiments show that the resulting changes to the model atmospheres is  marginal in the case of changing $V_t$ from 3.5 to 4.5 km/s.

The results of our fitting are shown in  Fig. \ref{sel1}. The accuracy of the abundance determination is not high ($\pm$ 0.5 dex), mainly due to the  low resolution of our spectra.
In the case of J052520, we obtained similar results from the blue and red spectral ranges. However, the red spectral range provides
less confident results. In the case of J052043, while fitting the model with the s-process elements enhanced, we reduced the abundances of Ti and Fe by -0.5 dex to get a better fit in the spectrum at 6200-6500 \AA. 

\begin{figure*}
\includegraphics[width=1.0\columnwidth]{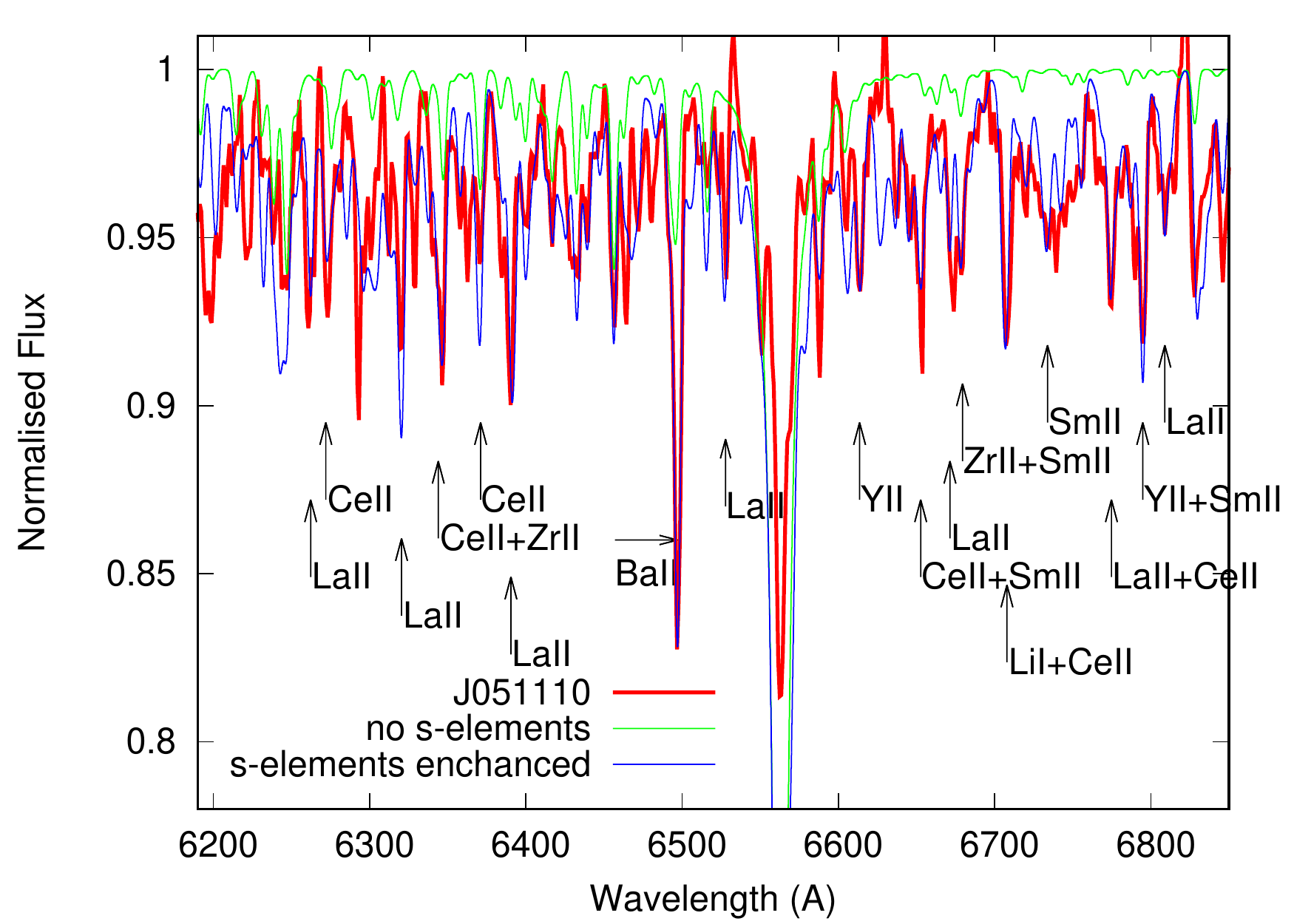}
\includegraphics[width=1.0\columnwidth]{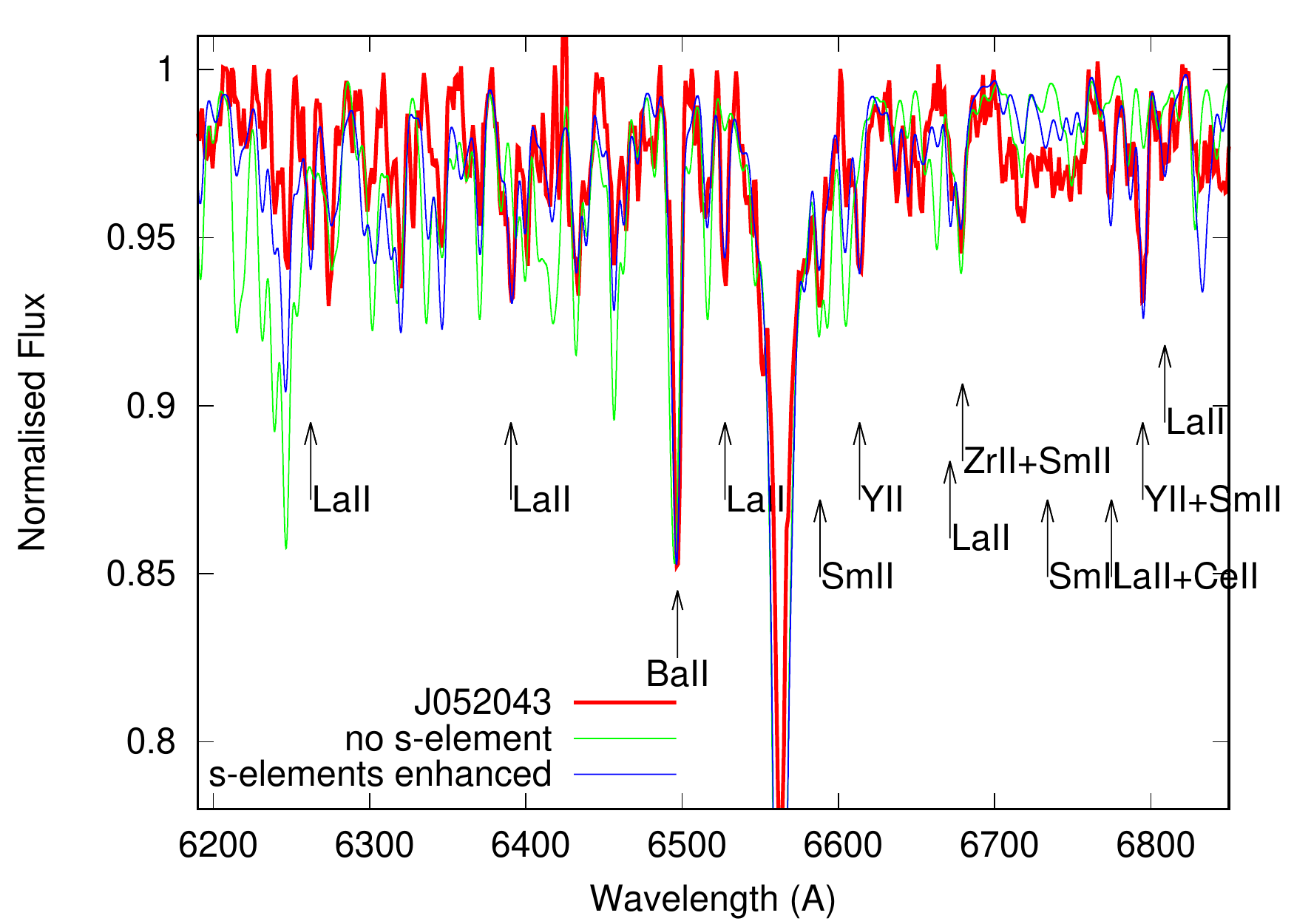}
\includegraphics[width=1.0\columnwidth]{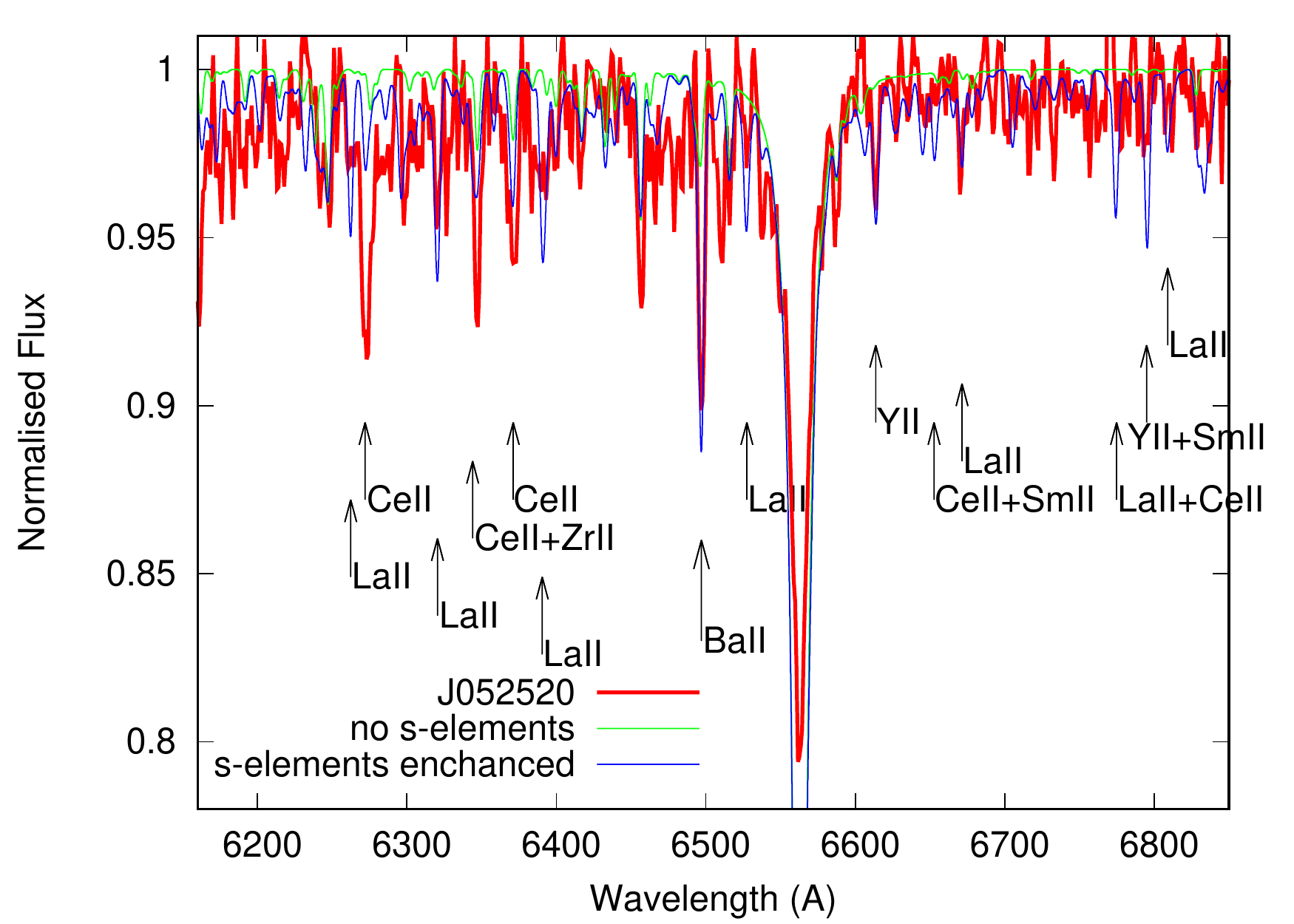}\hskip5mm
\includegraphics[width=1.0\columnwidth]{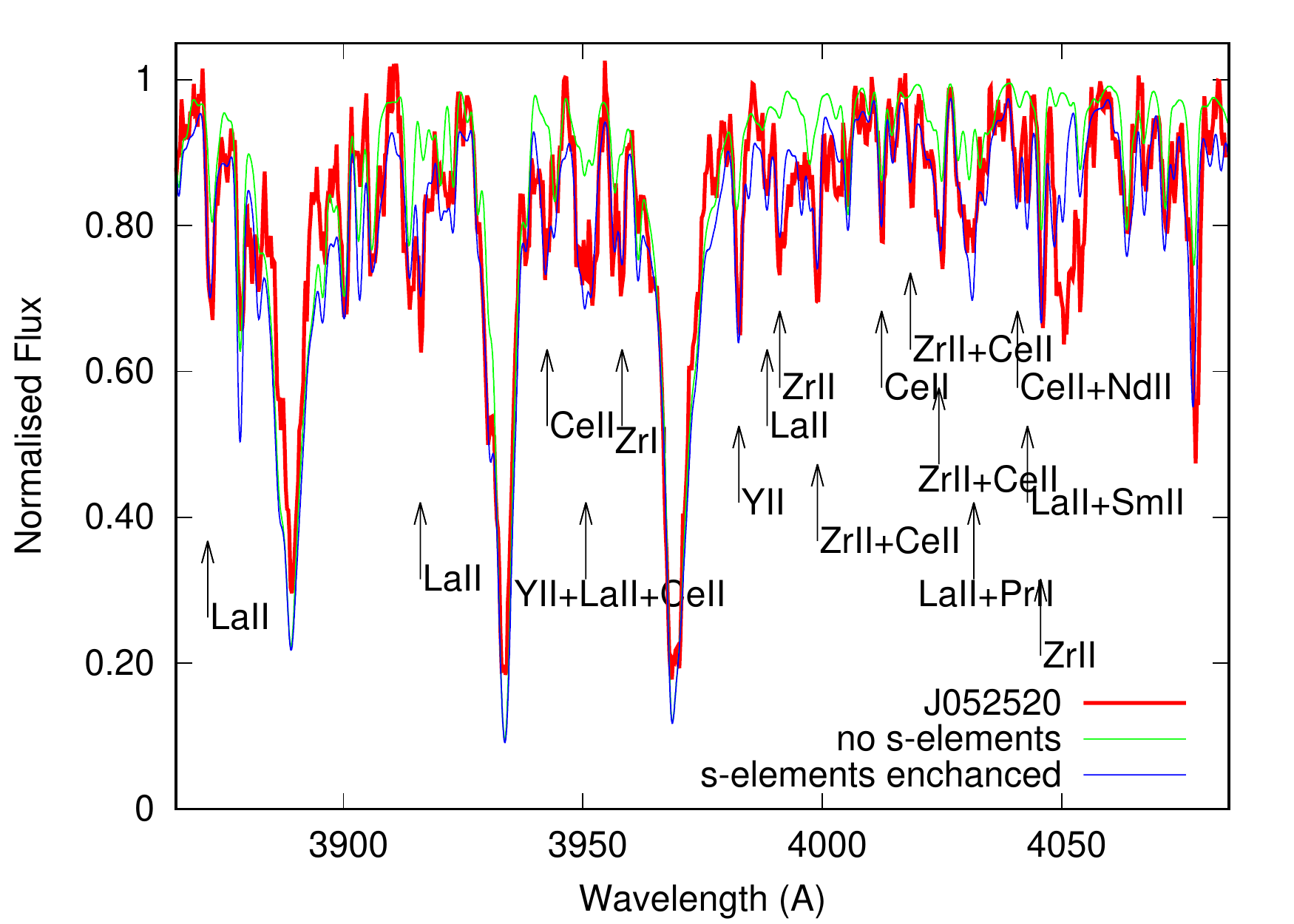} \\

\caption{Best fit of synthetic spectra with enhanced abundances of s-elements to the observed spectra of J051110 (top left), J052043 (top right), J052520 (bottom left), and blue spectrum of J052520 (bottom right). The fit with s-process elements enhanced to the spectrum of J052043 required a reduction in abundances of Ti and Fe of -0.5 dex }. 
\label{sel1}
\end{figure*}

Finally, we note that in the spectrum of J051110, we detected a strong  
absorption feature near 6708\,\AA.
\citet{2002A&A...395L..35R}, in a study of post-AGB stars, identified the feature at
6708\,\AA\ as a Ce\,{\sc ii} line $\lambda$ 6708.099\,\AA. However, to get the right fit to wavelength and strength of this feature in J051110 the contribution from Li\,{\sc i} 6707.8\,\AA\ line seems to be necessary.
In this spectral region, Ce\,{\sc ii} has three transitions at 6704.5\,\AA, 6706.1\,\AA,\ and 6708.1 \cite[see e.g.][]{2002A&A...395L..35R}, which in our resolution are blended into a single line, and a strong line at 6775\,\AA\ (see red line in Fig.\, \ref{lithium}). 

\begin{figure}
\resizebox{\hsize}{!}{\includegraphics{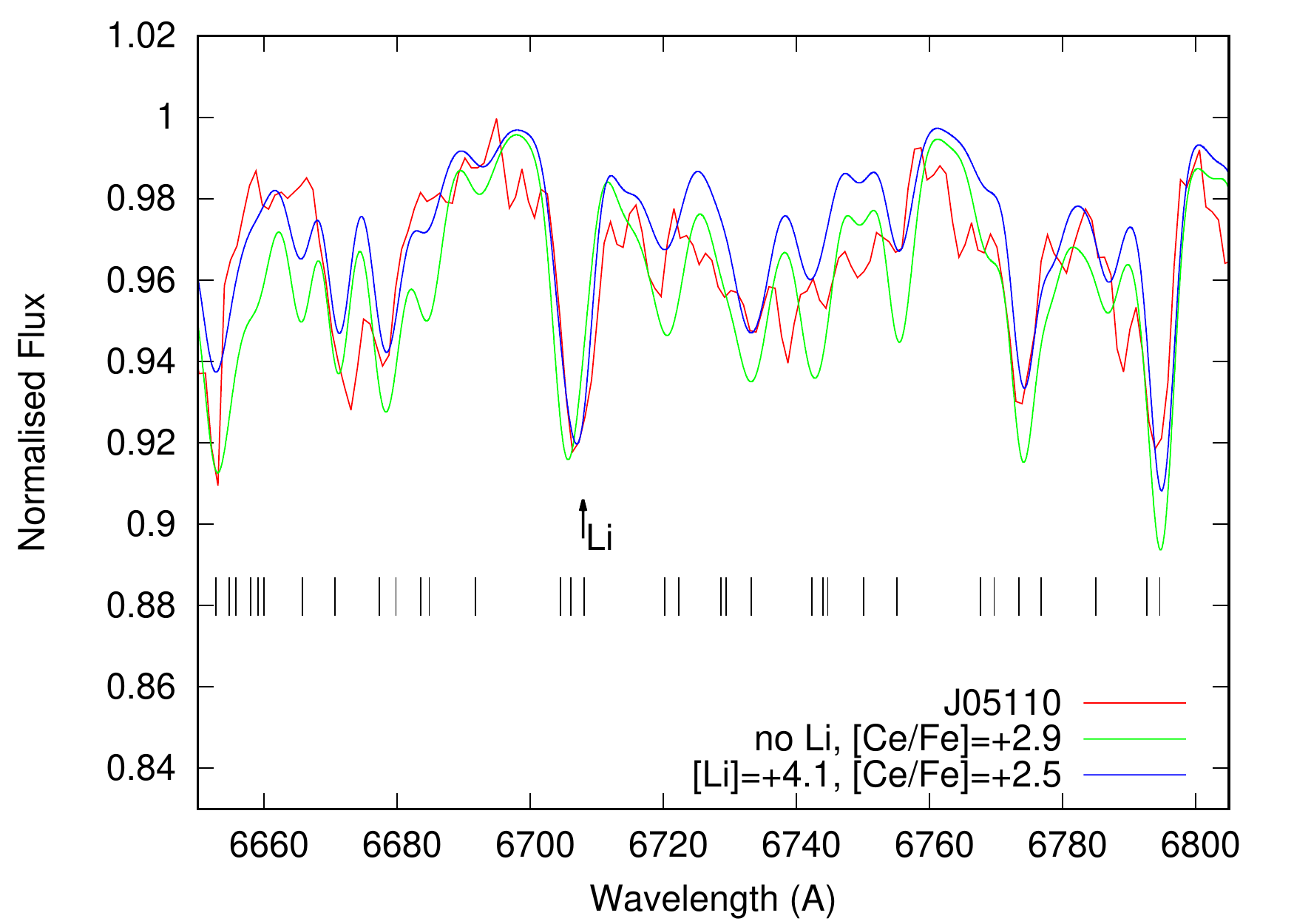}}
\caption{Observed spectrum of J05110 (red line) and two fits with enhancement of Ce and no Li (green line), and with both elements enhanced (blue line). See text for more details. The vertical lines show the positions of the Ce II lines.
}
\label{lithium}
\end{figure}

In order to fit the observed intensity of the 
6708\,\AA\ absorption and position of the 6775\,\AA\ line, we had to enhance the Ce abundance to [Ce/Fe]=+2.9 and to shift the theoretical spectrum by about 45 km/s (about 1\,\AA) to the right. We note that the resolution of our spectra is about 1500 and the position of each line is known with an accuracy of about 2\,\AA. However, this shift in wavelengths, while giving right position of the  Ce\,{\sc ii} 6775\,\AA\ line, does not seem to reproduce the position of the 6708\,\AA\ blend.
In addition, the assumed enhancement of Ce results in  absorption that is too strong at 6775\,\AA\ (see green line in Fig.\,\ref{lithium}). One of the obvious explanations for this discrepancy is that the lithium Li\,{\sc i} 6707.8\,\AA\ line contributes to the
observed blend at 6708\,\AA. This fit, with the large enhancement of [Li]=+4.1 and slightly reduced overabundance of Ce ([Ce/Fe]=+2.5), is shown  in Fig.\,\ref{lithium} (blue line). The match in position and strength for both features at 6708 and 6775\,\AA\ is much better now. However, the limited resolution of our spectrum does not allow us to perform a more detailed analysis. Spectra of much higher resolutions are necessary 
to confirm the presence of lithium in the atmosphere of J05110. Interestingly, the other stars in our sample do not show the 6708\,\AA\ feature.

\section{Discussion and conclusions}
\label{summ}

We observed fifteen optically bright post-AGB candidates and one spectroscopic standard in the MCs using the SALT telescope. The candidates were selected on the basis of the infrared colour-colour diagram [K]-24 versus [K]-[8], residing in the region where most of the known post-AGBs and PNe, as well as our HD evolutionary tracks, are located.
The spectra were obtained mainly in the red region (6160$-$9140\,\AA) and have relatively low resolution R$\sim$1500. Ten stars show emission lines or combined
emission and absorption lines in their spectra, one object has an essentially featureless spectrum, and four stars have absorption-line spectra and are likely post-AGB stars. 
We examined the long-slit spectra for diffuse background $\rm H \alpha$ emission; it is found in most of the emission-line objects, but in none of the likely post-AGB stars. The summary of the obtained results is presented in Tab.\,\ref{objects}.

Among the  emission-line objects, we classify three as young stellar objects (objects No.\,1, 2, and 8 according to the numbers from Table\,\ref{objlist}), two as possible Herbig Ae/Be stars (No.\,4 and 5), two as massive stars probably post-main sequence (No.\,6 and 14), one as a galaxy (No.\,10), one as a planetary nebula (No.\,12), and two as R CrB stars (No.\,13 and 15, the latter  classified as such for the first time).

For the four absorption-line objects (No.\, 3, 7, 9, and 11), we performed MK classification of the spectra and found them to range from B8 to F5 supergiants. We then fitted model atmospheres to the spectra to derive their physical parameters. The obtained model parameters confirm the post-AGB nature of the objects, and physical parameters were obtained for the first time for objects No.\,3 and 7. For three of them (No.\, 7, 9, and 11), we calculated the approximate abundances of the s-process elements, and for object No.\,11 we argued that lithium contributes to the absorption structure observed at about 6708\,\AA. All absorption-line objects are C-rich as they possess the characteristic 21 $\mu$m feature.  However, the C/O ratio
is poorly known in the atmospheres of our stars due to the lack of the appropriate absorption lines in their spectra.

\begin{figure}
\resizebox{\hsize}{!}{\includegraphics{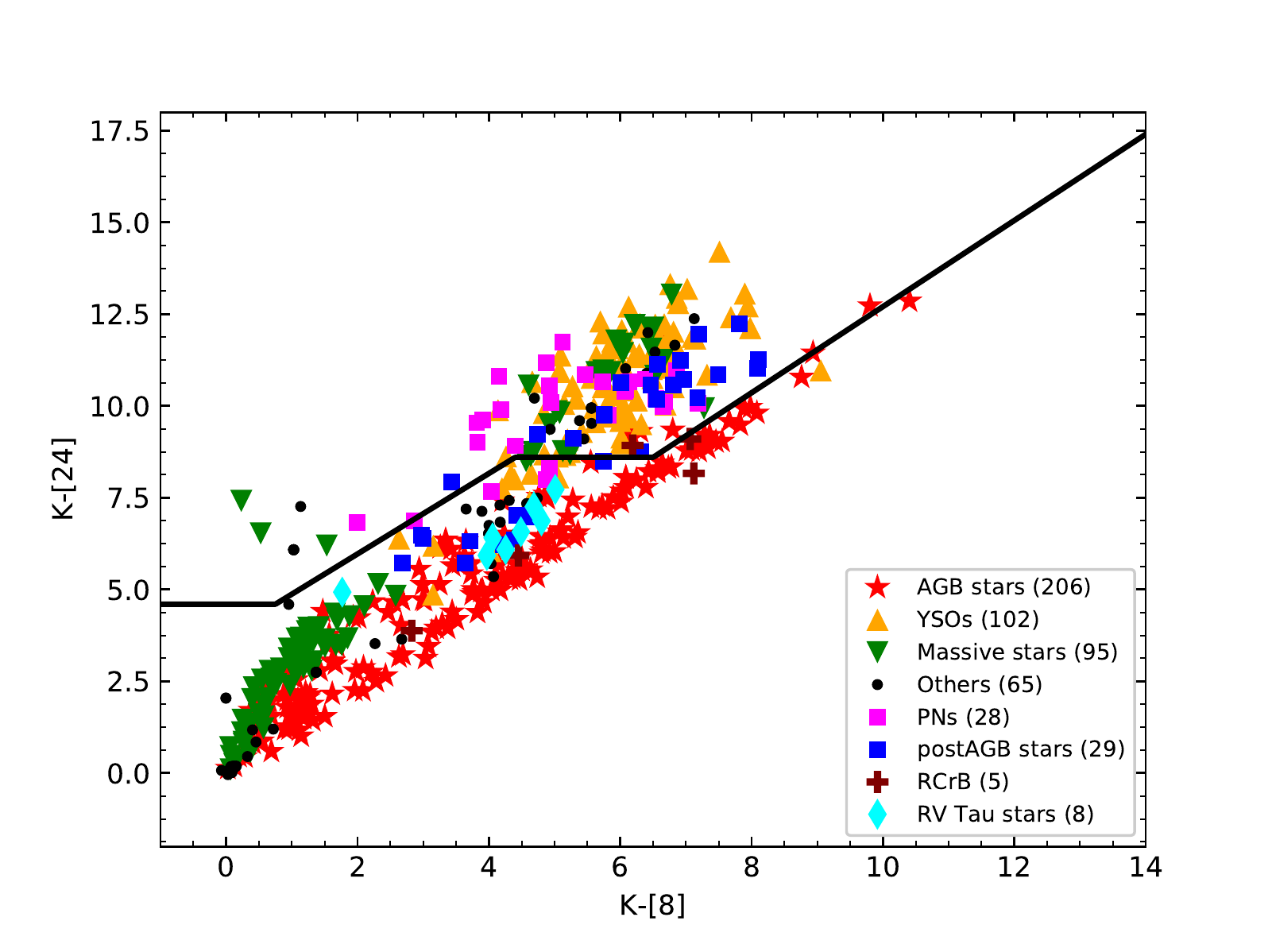}}
\caption{Colour-colour diagram used for selection of the post-AGB candidates (same as in Fig.\ref{pagb-select}). The main post-AGB candidate area is above the black lines. The grouped  class of objects (see text for details) from \citet{Jones:2017aa} are plotted with different symbols. The numbers in parentheses show the total number of each class of object plotted in this figure.}
\label{pagb-select-final}
\end{figure}

Among the 15 objects 
from our sample, we identified 7 evolved objects (four genuine post-AGB stars, two R CrB stars, and one PN), 7 young objects (three YSOs, two massive stars, and two Herbig Ae/Be stars), and one galaxy. Thus, we can conclude that the selected region on [K]-24 versus [K]-[8] colour-colour diagram is able to discriminate evolved objects at a success rate of about 50\%   (about 25\% for genuine post-AGB objects). The remaining objects are YSOs or massive post-main sequence stars taking into account their circumstellar dust emission. These statistics are valid only for
the brightest objects, which were selected in order to obtain optical spectroscopy with SALT at a reasonable S/N. We normally expect massive stars to be relatively rare in a general sample because they are relatively short-lived. However, by selecting the brightest objects our sample was biased toward the inclusion of massive evolved stars. 


It is possible to use an independent test to check the  usefulness of the area that we denoted  as 'post-AGB candidates area'. Recently, \citet{Jones:2017aa} classified about 1000 Spitzer spectra for about 800 sources in the Large Magellanic Cloud. These sources cover most  star evolutionary stages. Those with good photometry at K, [8], and [24] bands are plotted in Fig.\,\ref{pagb-select-final}. We counted together objects of different chemistry (C- and O-rich)  for AGB and post-AGB stars, and different classes of YSOs. The numbers in parentheses in the inset indicate the total number of sources in each class plotted in the figure. In the post-AGB candidate area (above the black line) there are altogether 172 objects: 84 YSOs; 26 massive stars; 14 others; 5 AGB stars; 23 PNe; 19 genuine post-AGB stars; and 1 RCrB star. As can be  seen, young objects compose more than 60\% of the total number of sources, while evolved stars less than 30\% of them. About 66\% (19/29) of the genuine post-AGB stars are located in the selected region, but they compose only about 11\% of the total number of sources above the black solid line in Fig.\ref{pagb-select-final}.

Thus the region we delineated  of the [K]-24 versus [K]-[8] colour-colour diagram as a `post-AGB candidate area' seems to select most of them, but it is populated by far more YSOs and other types of emission-line objects than it is by post-AGB stars.  Photometry at longer wavelength bands would allow us to better discriminate on the basis of colour, as it was in the case with the IRAS filters \citep[see e.g.][]{van-der-Veen:1988aa}. Until that is available, it appears that it is possible to use low-resolution spectra to eliminate the strong emission-line objects from consideration as post-AGB stars (but perhaps losing some PN). 
In addition, post-AGB objects of F--G spectral types are known to pulsate with periods from about 30 to about 150 days \citep[see e.g.][]{Hrivnak:2010aa,Hrivnak:2015aa}, so studies of light variability of the selected candidates may be helpful in identifying post-AGB objects.

\section{Acknowlegements}
The spectroscopic observations reported in this paper were obtained with the Southern African Large Telescope (SALT). 
Polish participation in SALT is funded by grant No. MNiSW
DIR/WK/2016/07. Our studies are partially supported by FP7 Project No:.
269193 ``Evolved stars: clues to the chemical evolution of galaxies".
This work was co-funded under the Marie Curie Actions of the European Commission (FP7-COFUND). RSz and NS acknowledge support from the NCN grant 2016/21/B/ST9/01626 and 2014/15/B/ST9/02111. 
MH thanks the Ministry of Science and Higher Education (MSHE) of the Republic of Poland for granting funds for the Polish contribution to the International LOFAR Telescope (MSHE decision no. DIR/WK/2016/2017/05-1) and for maintenance of the LOFAR PL-612 Baldy (MSHE decision no.~59/E-383/SPUB/SP/2019.1). 
BJH was supported by NSF AST 1413660.
This research has made use of the SIMBAD database, operated at CDS, Strasbourg, France.
We thank the anonymous Referee for his/her thorough review and highly appreciate the comments and suggestions, which significantly contributed to improving the quality of the publication.

\bibliographystyle{aa} 
\bibliography{postagb} 


\newpage
\appendix

\newpage
\section{Colour-magnitude diagrams.}
\label{appB}
The four colour-magnitude diagrams investigated by \cite{Blum:2006aa} are shown here with application to the selected sample of post-AGB candidates. In each figure we plotted different classes of LMC sources from \citet{Jones:2017aa}, which have good corresponding photometry. We counted together objects with different chemistry (C- and O-rich)  for AGB and post-AGB stars, and different classes of YSOs. In addition, we  overplotted positions of  self-consistent, time-dependent hydrodynamical (HD) radiative transfer calculations for gaseous dusty circumstellar shells around C-rich and O-rich stars in the final stages of their AGB/post-AGB evolution. These calculations are based on 
the evolutionary track for a star with an initial
mass of 3.0\,M$_{\odot}$, which, due to mass loss, is reduced to typical mass for central stars of post-AGB objects of 0.605\,M$_{\odot}$ \citep{Bloecker:1995fk} at the end of the AGB evolution, and single dust  grain size (a=0.05 $\mu$m) for both chemical compositions: amorphous carbon (AC) and astronomical silicates (ASil) \citep[see][for details] {Steffen:1998aa}. The HD evolutionary track for AC dust is shown by red/orange dots (C-AGB Track/C-PAGB Track in the  insets), while for ASil dust by blue/light blue dots (O-AGB Track/O-PAGB Track) for the AGB/post-AGB phase, respectively. 
The HD models are plotted in equidistant intervals in time (100 years for AGB and 0.3 year for post-AGB), so the density of the points is a direct measure of the probability of finding various objects  in different parts of the diagram. The last 200 years during AGB (the quick transition to post-AGB phase) are plotted each 20 years. The AGB phase covers 350\,000 years, while the post-AGB only about 1\,000 years. The confirmed post-AGB objects are indicated by the filled circles.

\begin{figure*}
\resizebox{\hsize}{!}{\includegraphics{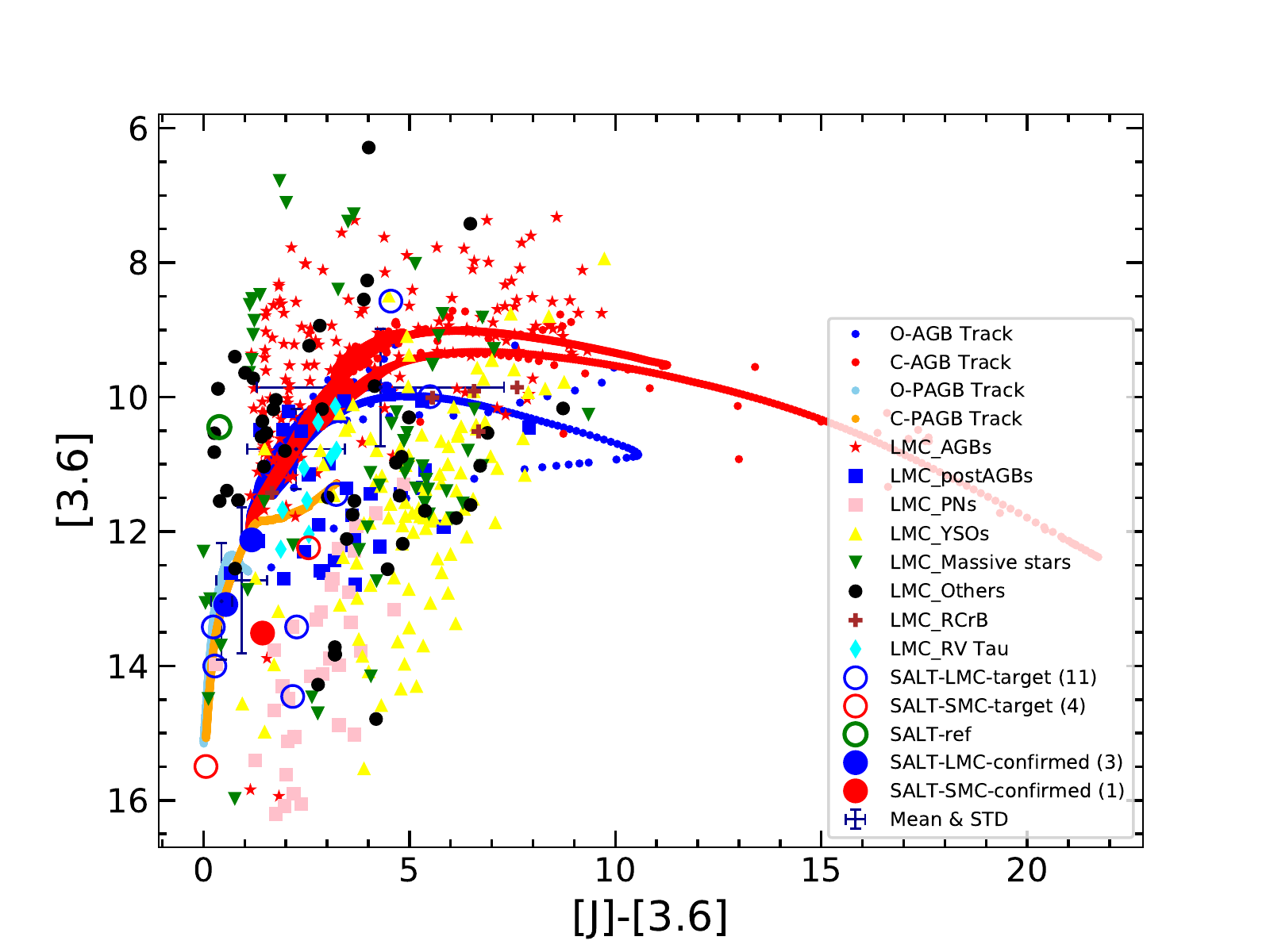}}
\caption{Colour-magnitude diagram [3.6] vs [J]-[3.6]. The selected targets for SALT observations are indicated by circles: black (LMC), red (SMC), and green (SK\,105). The confirmed post-AGB objects (No. 3, 7, and 11) are plotted as   filled circles and are located within 0$<$[J]-[3.6]$<2$ and 12>[3.6]>14, which may be compared with Fig.3 of \cite{Blum:2006aa}. Object No. 9 has no J photometry. }
\label{pagb-select-final-1}
\end{figure*}

\begin{figure*}
\resizebox{\hsize}{!}{\includegraphics{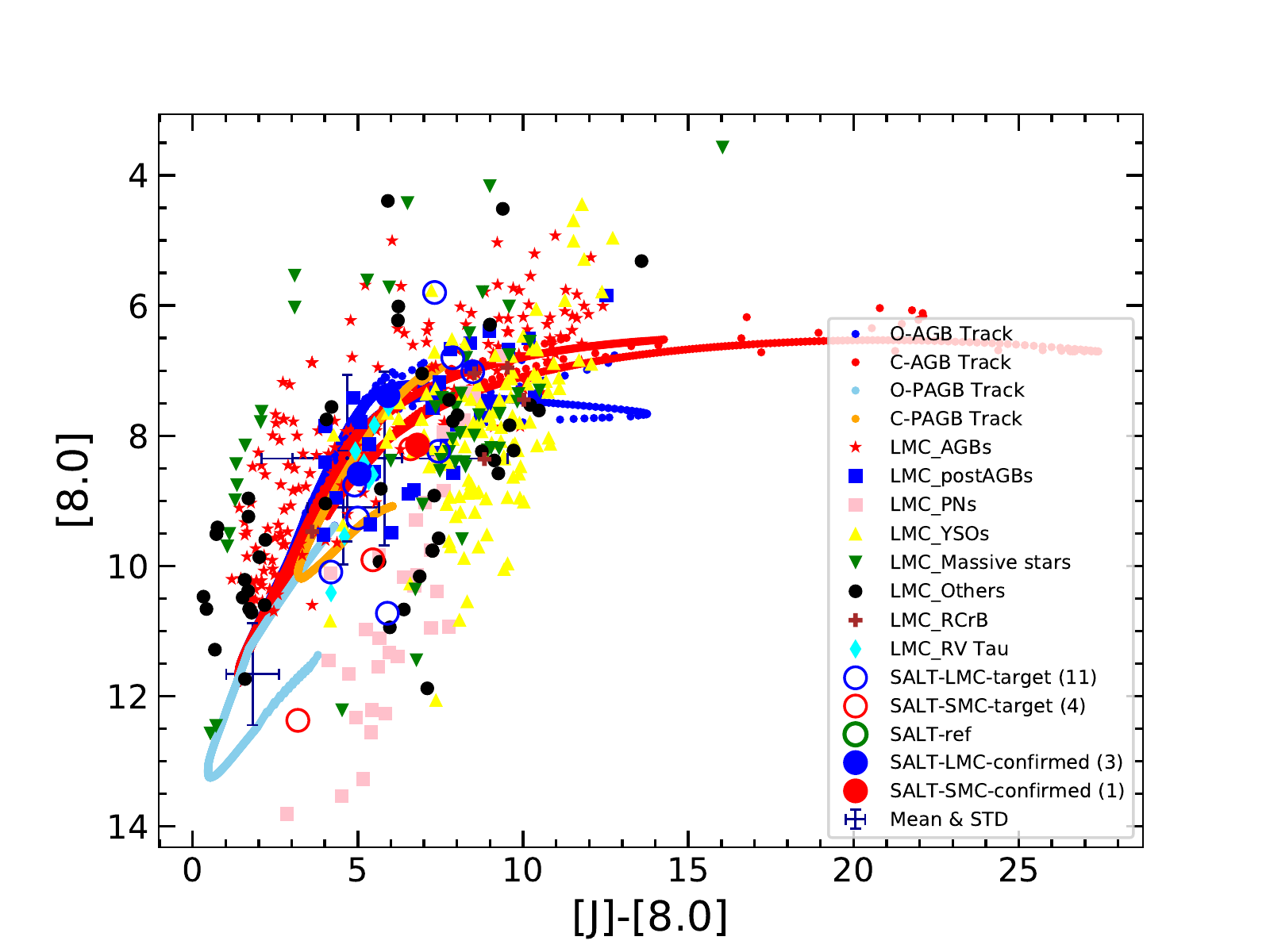}}
\caption{Colour-magnitude diagram [8.0] vs [J]-[8.0]. The selected targets for SALT observations are indicated by circles: black (LMC), red (SMC), and green (SK\,105). The confirmed post-AGB objects (No. 3, 7, and 11)  are plotted as   filled circles and are located within 5$<$[J]-[8.0]$<7$ and 7>[3.6]>9, which is located within the extreme AGB star region in Fig.4 of \cite{Blum:2006aa}. Object No. 9 has no J photometry.}
\label{pagb-select-final-2}
\end{figure*}

\begin{figure*}
\resizebox{\hsize}{!}{\includegraphics{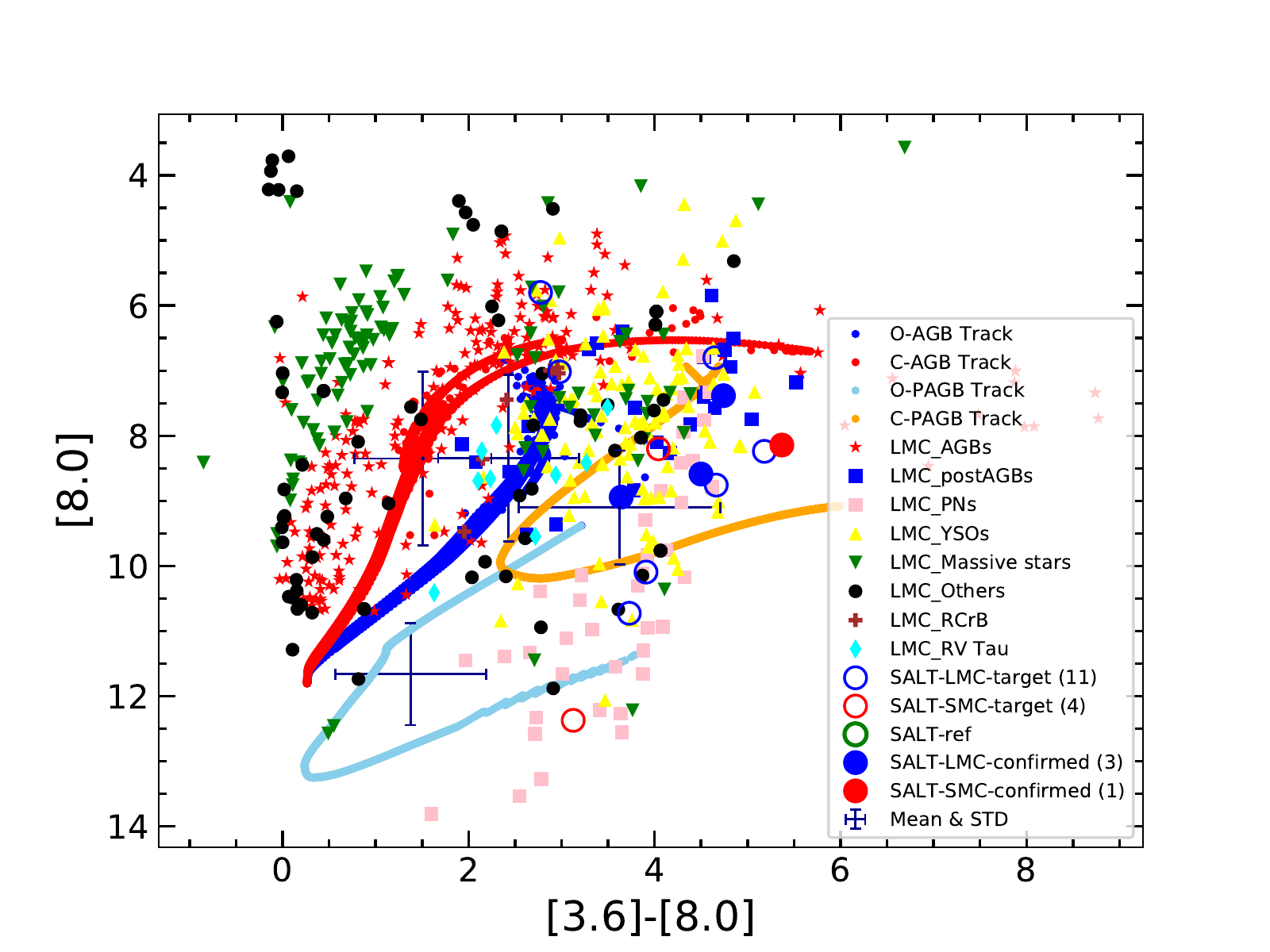}}
\caption{Colour-magnitude diagram [8.0] vs [3.6]-[8.0]. The selected targets for SALT observations are indicated by circles: black (LMC), red (SMC), and green (SK\,105). The confirmed post-AGB objects (No. 3, 7, 9, and 11)  are plotted as filled circles and are located within 3$<$[3.6]-[8.0]$<6$ and 7>[8.0]>9, which can be compared with Fig.5 of \cite{Blum:2006aa}.}
\label{pagb-select-final-3}
\end{figure*}

\begin{figure*}
\resizebox{\hsize}{!}{\includegraphics{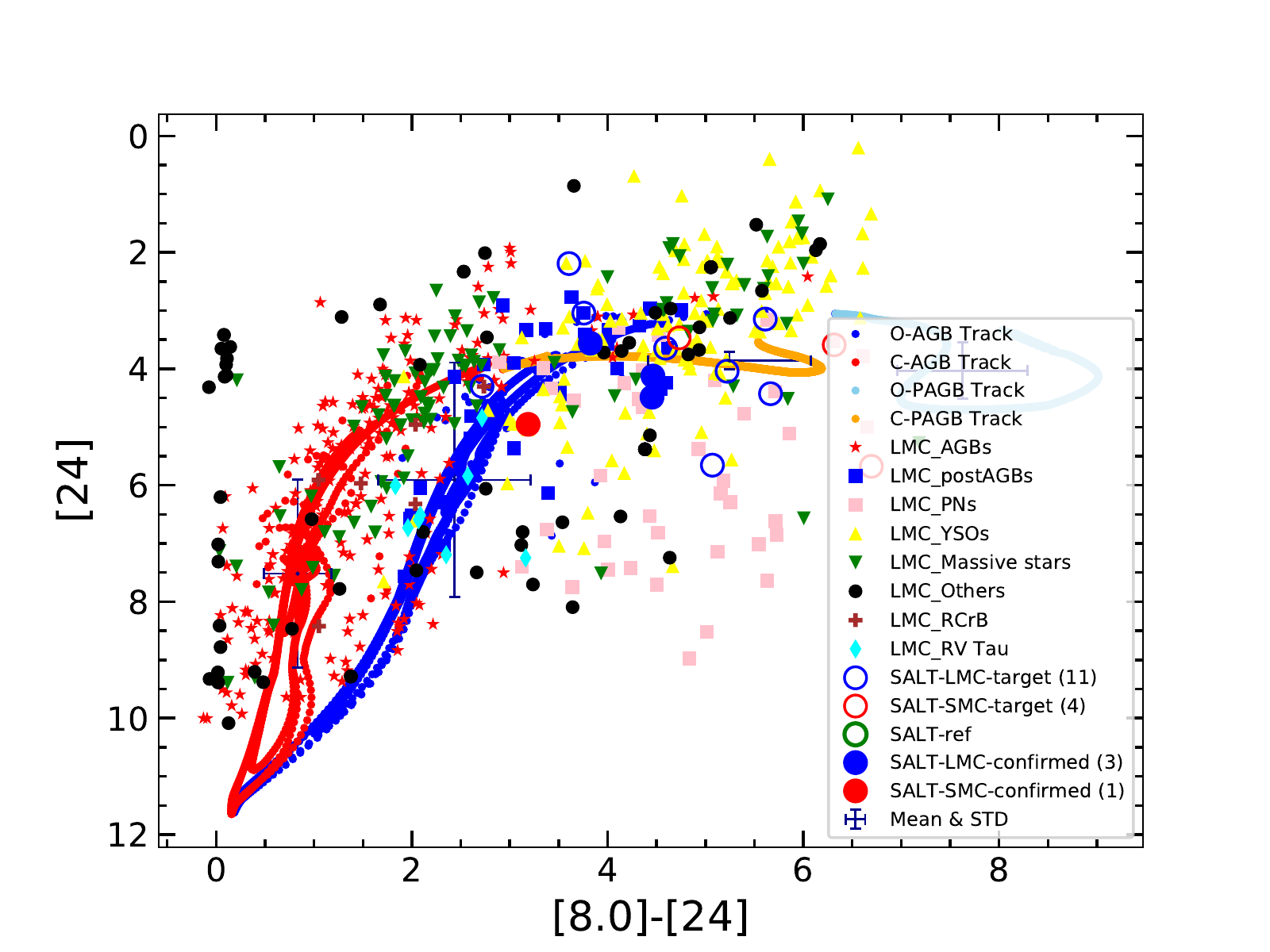}}
\caption{Colour-magnitude diagram [24.0] vs [8.0]-[24]. The selected targets for SALT observations are indicated by circles: black (LMC), red (SMC), and green (SK\,105). The confirmed post-AGB objects (No. 3, 7, 9, and 11) are plotted as filled circles and are located within 3$<$[8.0]-[24]$<5$ and 3>[24]>5, which can be compared with Fig.6 of \cite{Blum:2006aa}.}
\label{pagb-select-final-4}
\end{figure*}

\end{document}